\documentclass[a4paper,fleqn,usenatbib]{mnras}

\usepackage[T1]{fontenc}
\usepackage{ae,aecompl}

\usepackage{graphicx}	
\usepackage{amsmath}	
\usepackage{amssymb}	

\newcommand{\MS}{\ifmmode{\,}\else\thinspace\fi{\rm M}\ifmmode_{\odot}\else$_{\odot}$\fi}
\newcommand{\LS}{\ifmmode{\,}\else\thinspace\fi{\rm L}\ifmmode_{\odot}\else$_{\odot}$\fi}
\newcommand{\RS}{\ifmmode{\,}\else\thinspace\fi{\rm R}\ifmmode_{\odot}\else$_{\odot}$\fi}
\newcommand{\ff}{\ifmmode \nu_{\rm F}\else$\nu_{\rm F}$\fi}
\newcommand{\fo}{\ifmmode \nu_{\rm 1O}\else$\nu_{\rm 1O}$\fi}
\newcommand{\fx}{\ifmmode \nu_{\rm x}\else$\nu_{\rm x}$\fi}
\newcommand{\ax}{\ifmmode A_{\rm x}\else$A_{\rm x}$\fi}
\newcommand{\Po}{\ifmmode P_{\rm 1O}\else$P_{\rm 1O}$\fi}
\newcommand{\Pf}{\ifmmode P_{\rm F}\else$P_{\rm F}$\fi}
\newcommand{\Px}{\ifmmode P_{\rm x}\else$P_{\rm x}$\fi}
\newcommand{\pxpo}{\ifmmode P_{\rm x}/P_{\rm 1O}\else$P_{\rm x}/P_{\rm 1O}$\fi}
\newcommand{\pxpf}{\ifmmode P_{\rm x}/P_{\rm F}\else$P_{\rm x}/P_{\rm F}$\fi}

\title[Peculiar double-periodic pulsation in RR~Lyr stars]{Peculiar double-periodic pulsation in RR~Lyrae stars of the OGLE collection. II. Short-period stars with dominant radial fundamental mode.}
\author[Prudil et al.]{
Z. Prudil$^{1,2}$\thanks{E-mail: prudilz@ari.uni-heidelberg.de}, R. Smolec$^{3}$, M. Skarka$^{4}$, H. Netzel$^{5}$
\\
$^{1}$ Astronomisches Rechen-Institut, Zentrum f{\"u}r Astronomie der Universit{\"a}t at Heidelberg, M{\"o}nchhofstr. 12-14,\\ \,\,\,\,\,69120 Heidelberg, Germany\\
$^{2}$ Department of Theoretical Physics and Astrophysics, Masaryk University, Kotl\'a\v{r}sk\'a 2, 611 37 Brno, Czech Republic\\
$^{3}$ Nicolaus Copernicus Astronomical Center, Polish Academy of Sciences, ul. Bartycka 18, 00-716 Warszawa, Poland\\
$^{4}$ Konkoly Observatory, MTA Research Centre for Astronomy and Earth Sciences, Konkoly Thege \'{u}t 15-17, H-1121 Budapest, Hungary \\
$^{5}$ Warsaw University Observatory, Al. Ujazdowskie 4, 00--478 Warszawa, Poland\\
}

\date{Accepted . Received ; in original form }

\pubyear{2016}

\begin{document}
\label{firstpage}
\pagerange{\pageref{firstpage}--\pageref{lastpage}}
\maketitle

\begin{abstract}
We report the discovery of a new group of double-periodic stars in the OGLE Galactic bulge photometry. In 38 stars identified as fundamental mode RR~Lyrae and 4 classified as the first-overtone RR~Lyrae, we detected additional shorter periodicity. Periods of the dominant variability in the newly discovered group are $0.28<P_{\rm D}<0.41$\,days. Period ratios $(0.68 - 0.72)$ are smaller than the period ratios of the Galactic bulge RRd stars. The typical amplitude ratio (of the additional to the dominant periodicity) is 20\,\% for the stars identified as fundamental mode RR~Lyrae and 50\,\% for stars classified as the first-overtone RR~Lyrae. Ten stars from our sample exhibit equidistant peaks in the frequency spectrum, that suggest the Blazhko-type modulation of the main pulsation frequency and/or the additional periodicity. The Fourier coefficients $R_{\mathrm{21}}$ and $R_{\mathrm{31}}$ are one of the lowest among fundamental mode RR Lyrae stars, but among the highest for the first-overtone pulsators. For the phase Fourier coefficients $\varphi_{\mathrm{21}}$ and $\varphi_{\mathrm{31}}$, our stars lie in between RRab and RRc stars. Discussed stars were compared with the radial, linear pulsation models. Their position in the Petersen diagram cannot be reproduced assuming that two radial modes are excited and their physical parameters are similar to that characteristic for RR~Lyrae stars. The non-radial mode scenario also faces difficulties. We conclude that the dominant variability is most likely due to pulsation in the radial fundamental mode including stars classified as the first overtone mode pulsators. At this point, we cannot explain the nature of the additional periodicity. Even more, classification of the stars as RR~Lyrae should be treated as tentative.
\end{abstract}




\section{Introduction}\label{sec:intro}

Recent studies of RR Lyrae stars based on ultra-precise, quasi-continuous space observations ({\it MOST}, {\it CoRoT} and {\it Kepler}) and top-quality long-term ground-based measurements represented by the Optical Gravitational Lensing Experiment \citep[OGLE][]{ogleIII,ogleIV} completely changed our view on these stars revealing the true complexity of their pulsational behaviour. Some of the RR Lyrae stars, which were previously considered to pulsate purely radially in the fundamental (F mode, RRab), first-overtone (1O mode, RRc), or in both modes simultaneously (F/1O modes, RRd), were found to show additional pulsational modes with low amplitude. 

\citet{aqleo} discovered AQ Leo (RRd star) to be the first example of the most numerous group of stars showing additional variability (lower index `A') with period $P_{\rm A}\in(0.6,0.64)P_{\rm1O}$ in {\it MOST} data. This peculiar periodicity, with amplitude in the mmag regime, was detected in photometry collected by space telescopes {\it CoRoT} and {\it Kepler} \citep[see e.g. references][]{szabo_corot, pamsm15}, but mainly in the OGLE Galactic bulge data where more than 260 such stars of both RRc and RRd types were identified \citep{netzel1,netzel3}. The most plausible recent explanation for the peculiar $P_{\rm A}$ variability counts with the non-radial modes trapped in the outer part of the envelope \citep{wd16}. Beside these so called 0.61-stars, first-overtone pulsators also show additional periodicity with $P_{\rm1O}/P_{\rm A}\sim0.686$ \citep{netzel2,nspta}, which has not yet been explained. Recently another new subclass of double-mode RR Lyraes had been identified \citep{Soszysky2016MNRAS}. These 22 anomalous RRd stars discovered in the Magellanic system, have period ratios in the range $0.725 - 0.738$. In most cases, the fundamental mode is the dominant periodicity among these stars. RRd stars with atypical period ratios detected earlier in the Galactic bulge \citep{Smolec2015} and in M3 \citep{Jurcsik2014} can also be classified as anomalous RRd stars.

In the first paper of the series \citep{smolec2016} we report the discovery of another group of fundamental mode RR Lyrae stars showing additional mode with $P_{\rm A}/P_{\rm F}\in(0.70;0.75)$\footnote{A few stars in this period-ratio range were identified in {\it Kepler/K2} by \citet{molnar2016}.} and with long fundamental mode period. This period ratio and pulsational models suggest that these stars could be an extreme case of RRd stars. 

In this paper, we present results on 38 stars catalogued by OGLE as RRab stars and on 4 stars identified as RRc variables with pulsation periods $0.28<P_{\rm D}<0.41$\,days. Our stars seem to be adjacent to the sequence formed by the Galactic bulge classical RRd stars in the Petersen diagram with shorter periods and lower period ratios $(0.68 - 0.72)$ (see Fig.~\ref{fig:petersen}). In the Sect.\,\ref{sec:methods} we provide a short summary about used methods. The Sect.\,\ref{sec:results} contains a comparison of our newly identified double-periodic stars with other RR~Lyrae subtypes together with brief notes on selected stars. In addition, we examine the properties of our newly discovered group and discuss several stars in detail. In section \ref{sec:discussion} we discuss the pulsation nature of the newly discovered group. Section \ref{sec:Conclusions} summarises our results. The appendix contains a short list of stars with an additional variability, that do not fall into our group but were detected during the conducted analysis.

\section{Data analysis}\label{sec:methods}

After identification of the first long-period, double-periodic RR Lyrae stars \citep{smolec2016}, we searched for additional double-periodic stars with period ratio close to 0.7 among all RRab and RRc stars observed by OGLE-IV\footnote{For further analysis we used combined OGLE-IV and OGLE-III data sets wherever it was possible.} in the Galactic bulge (36\,190 stars in total). We used automatic procedure \textsc{pysca} \citep{Herzberg2014} employing nonlinear-least squares methods and the Lomb-Scargle algorithm for identification of the suspicious periodicity. We looked for an additional signal (with ${\rm S/N}>4.0$) in range from $\nu_{\mathrm{D}}+0.3$c/d up to $2\nu_{\mathrm{D}}-0.3$c/d to avoid possible contamination with stars exhibiting Blazhko effect. If we found such signal, the star was sassigned to a candidate group and analysed manually in the same way as in our previous study \cite{smolec2016}: 

We followed a standard consecutive pre-whitening technique (fitting sine series). We considered only well resolved significant peaks (${\rm S/N}>4.0$) with separation larger than $2/\Delta T$, where $\Delta T$ is data length. If necessary, the outliers were omitted (4$\sigma$ clipping) and secular trends were removed using low-order polynomials or spline functions. We also analyzed both magnitude and flux-scaled data to avoid confusion of possible artificial peaks with the combination peaks. Combination peaks, if present in the flux-scaled data, are a convincing evidence that two detected frequencies are intrinsic to the star. Their presence is essential for marking the star as a candidate. The stars without combination peaks were accepted only if they shared similar properties with other stars within our group (period ratio, shape of the light curve). For details of all steps of the analysis see sect. 2 in \citep{smolec2016}. 

In addition, to the automatic procedure, we manually analysed all stars located in the same region of period-amplitude diagram as newly discovered double-periodic stars\footnote{$P_{\rm{D}}=\left \langle 0.28; 0.42 \right \rangle$ days and $A_{I}=\left \langle 0.35; 0.65 \right \rangle$ mag.}, to ensure completness of our sample.

\section{Results}\label{sec:results}

\begin{figure}
\centering
\resizebox{\hsize}{!}{\includegraphics{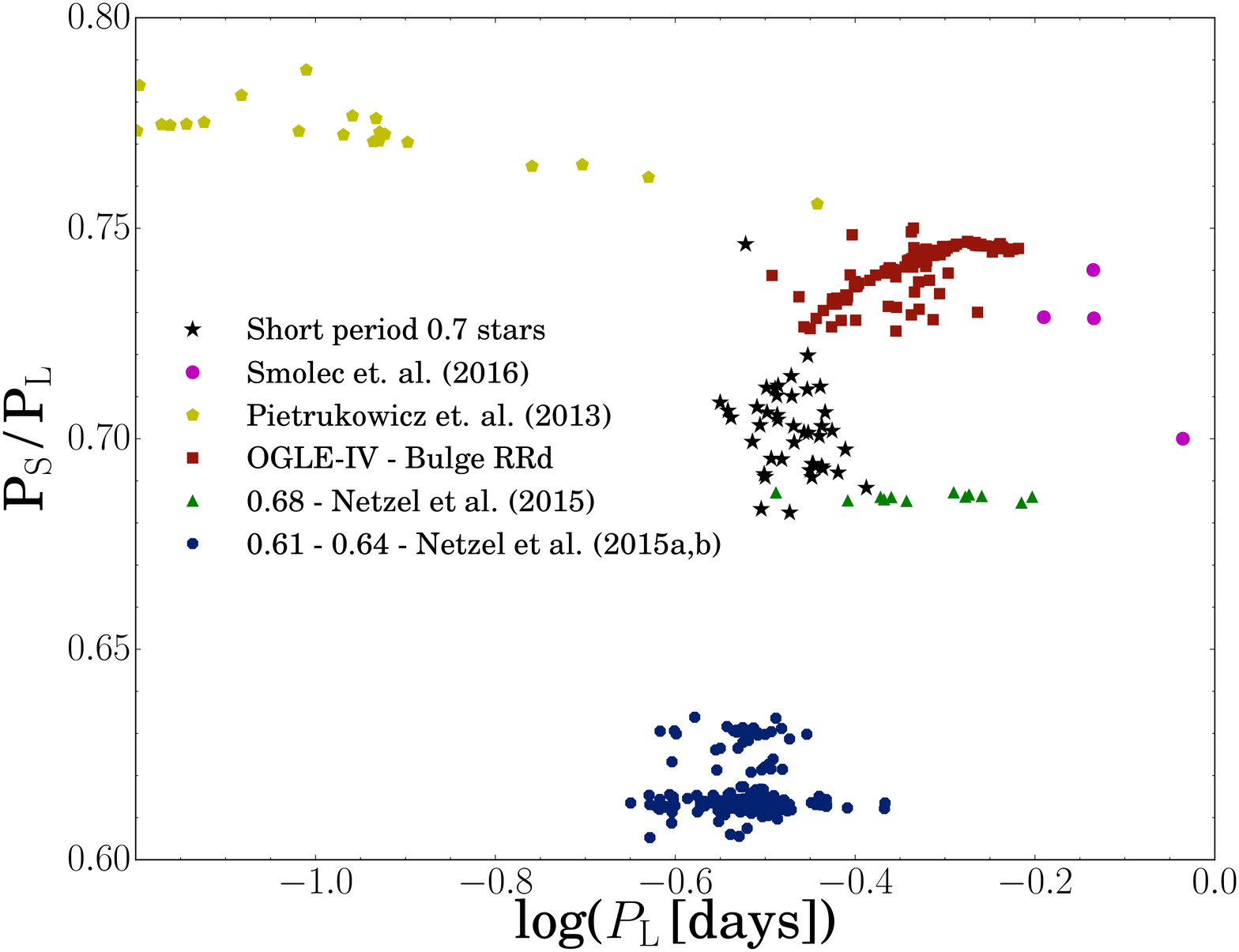}}
\resizebox{\hsize}{!}{\includegraphics{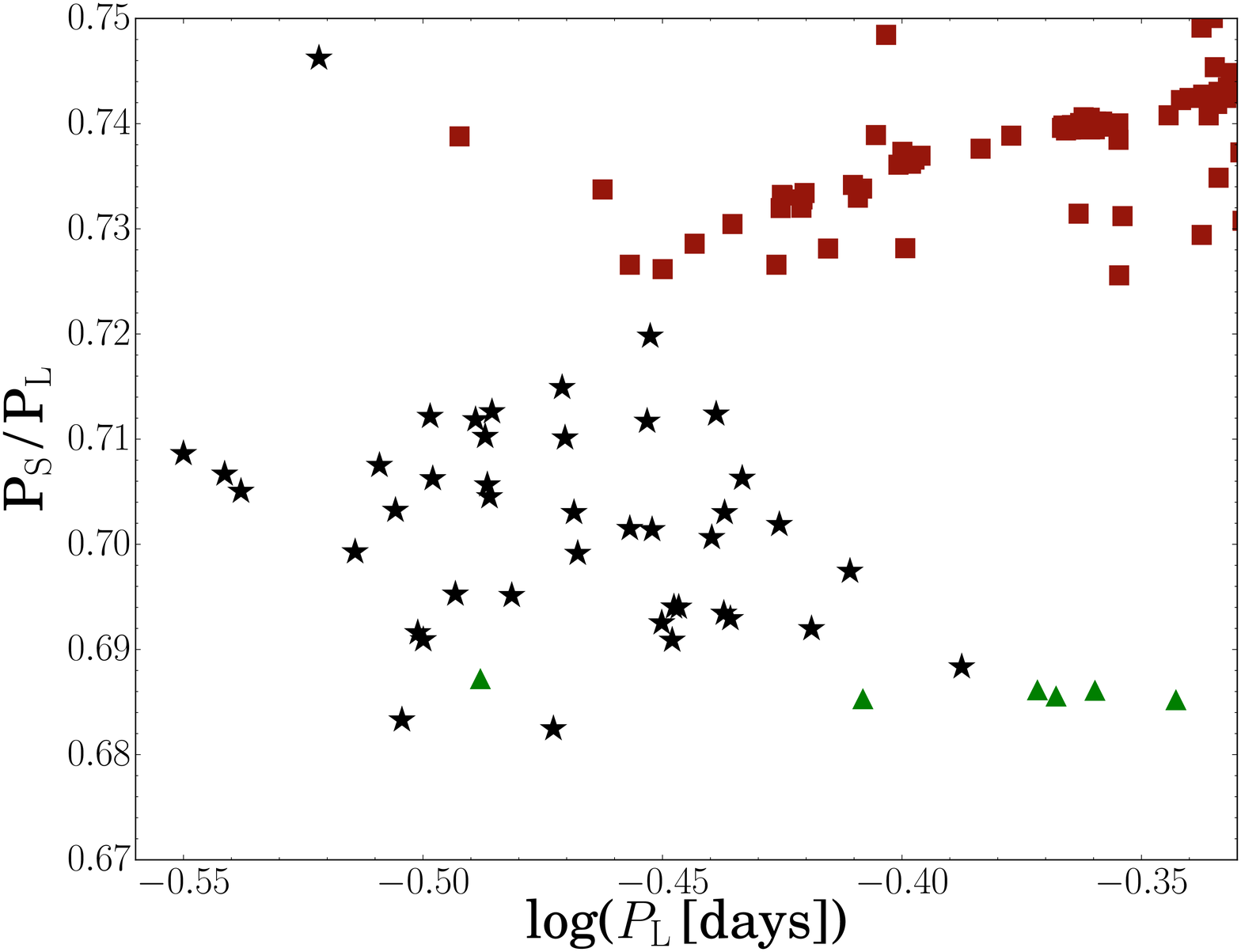}}
\caption{The Petersen diagram for newly discovered double-periodic stars (black stars). $P_{\mathrm{L}}$ and $P_{\mathrm{S}}$ are longer and shorter period, respectively. The top panel displays several groups of double-periodic stars in comparison with our group. The magenta circles represent double-periodic stars found by  \citet{smolec2016}. The yellow pentagons stand for several F+1O HADS stars found in the Galactic disk by \citet{Pietrukowicz2013}. In addition, green triangles and blue circles correspond to the stars found by \citet{netzel2} and \citet{netzel1,netzel3}. The bottom panel provides narrow context between OGLE-IV Galactic bulge RRd stars (red squares in both plots) and our group.}
\label{fig:petersen}
\end{figure}

From a total sample of 36\,190 single periodic RR~Lyrae stars in the OGLE-IV Galactic bulge catalog, we report a discovery of extra periodicity in 42 stars. In 14 of our stars, additional periodicity was reported by \cite{ogleIV_rrl_blg,ogleIII_rrl_blg}, but not discussed at all. It was only mentioned in the \textit{remarks} file available at OGLE ftp archive\footnote{\textit{ftp://ftp.astrouw.edu.pl/ogle/ogle4/OCVS/blg/rrlyr/remarks.txt}}. From this sample, 38 stars have been classified by \cite{ogleIV_rrl_blg} as the fundamental-mode stars, and 4 stars as the first-overtone pulsators. In all stars, the period ratio of the two periodicities falls in the $\approx \left(0.68-0.72 \right)$ range. The extra variability always has a shorter period and a lower amplitude than the dominant mode. In addition, in OGLE-BLG-RRLYR-14135 one more periodicity was detected with period ratio 0.74623. 
 
\begin{figure}
\centering
\resizebox{\hsize}{!}{\includegraphics{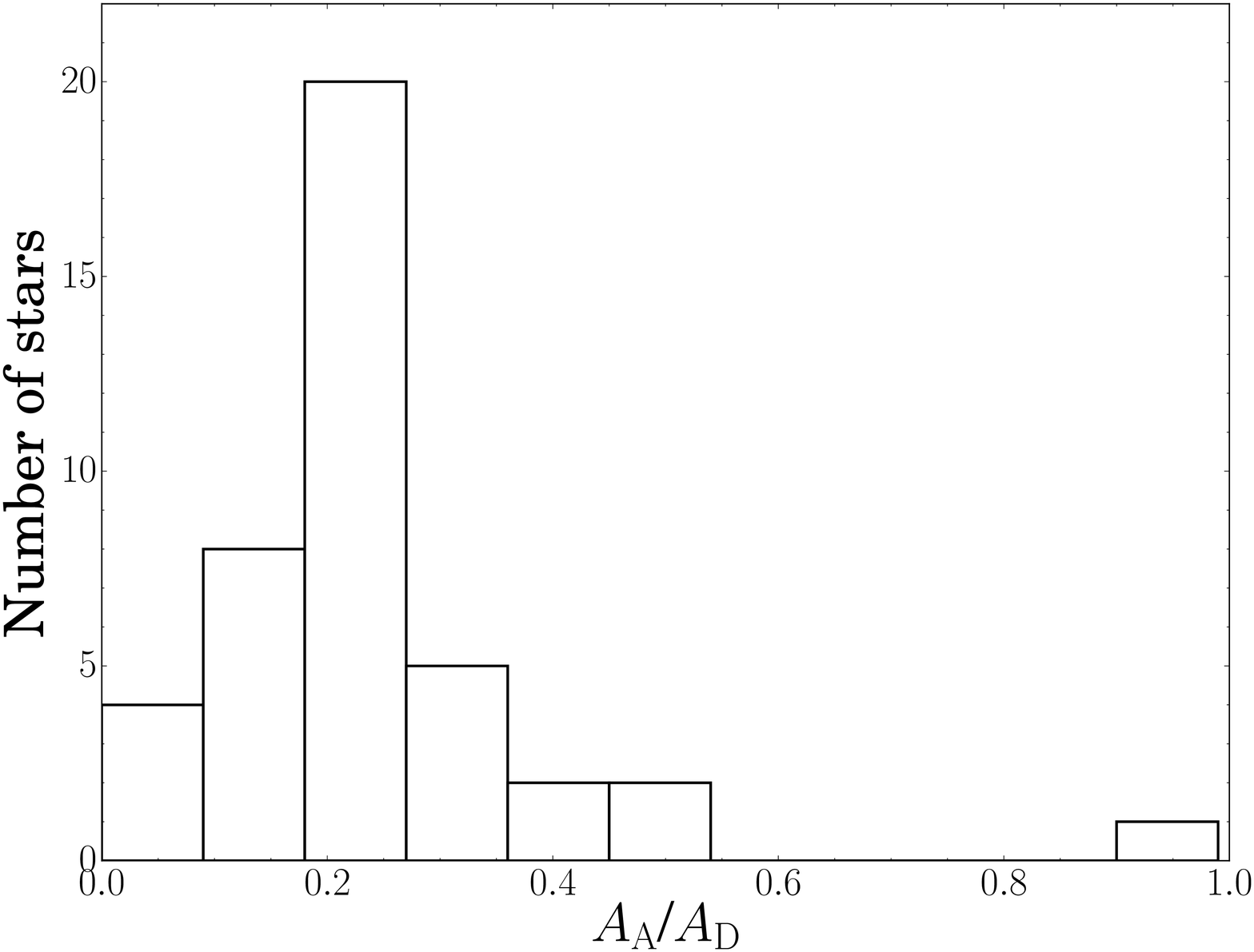}} \hspace{1.0cm}
\resizebox{\hsize}{!}{\includegraphics{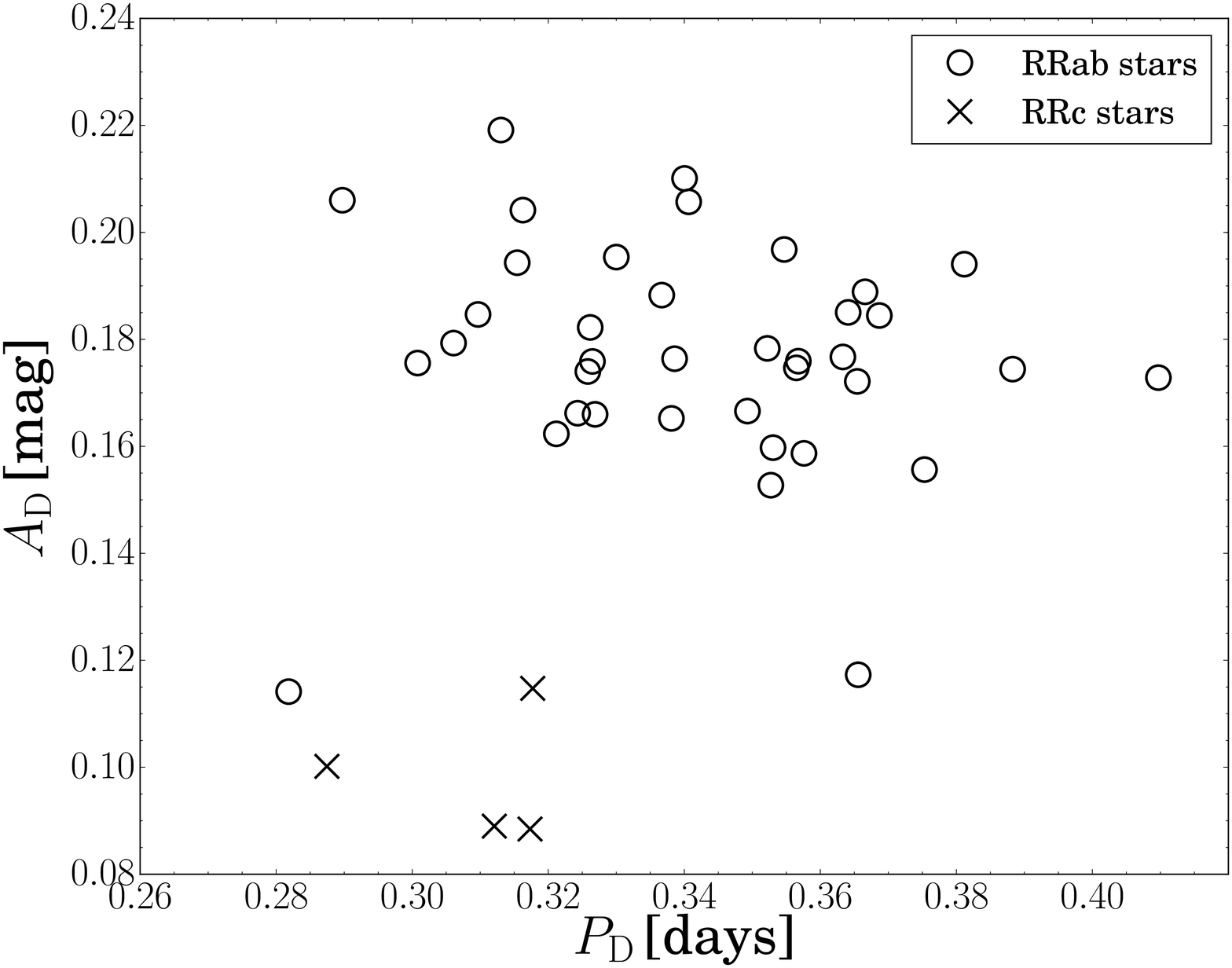}}
\caption{The distribution of amplitude ratio and the period-amplitude diagram for newly discovered double-periodic stars. The top-panel shows a histogram of amplitude ratios between dominant and additional periodicity. The bottom panel displays period-amplitude diagram for the dominant periodicity among discovered double-periodic stars with regard to their OGLE classification.}
\label{fig:ADPD}
\end{figure}

In the top plot of Fig.~\ref{fig:petersen}, we compare known double-periodic RR~Lyrae stars from our Galaxy, with a newly discovered group (black stars). Several double-periodic stars, identified as RRab, have a similar period ratio as the double-periodic pulsators found by \citet[][ marked with green triangles]{netzel2}. The latter stars are very different, however. The light curve of the dominant variability in these stars is among the most typical for the first overtone RR Lyrae stars and additional variability is always of a longer period with significantly lower amplitude (around 4\,\% of the first overtone mode amplitude). Furthermore, these stars form a well defined and sharp sequence in the Petersen diagram, while our stars do not. The newly discovered stars share some similarities with double-periodic stars found by \cite{smolec2016} (magenta points in the Petersen diagram). \cite{smolec2016} identified additional periodicity among fundamental mode stars, where the extra mode has a very low amplitude (down to 8.5\,\% of the fundamental mode). Our group, on the other hand, stands on the other side of the period spectrum for RR~Lyraes. Furthermore, stars from our group have somewhat larger amplitudes for the additional variability (around 25\,\% of the dominant periodicity, see the top panel of Fig.~\ref{fig:ADPD}). The stars with additional variability are similar to RRd stars (red squares), but the period ratio is lower around 0.7 (see bottom panel of Fig.~~\ref{fig:petersen}). 

The only outlier in the Petersen diagram is OGLE-BLG-RRLYR-14135. This star was identified as a fundamental mode pulsator and two extra periodicities were found. The first additional variability has period $P_{\mathrm{A_{\rm 1}}}=0.2244660(9)$\,days and period ratio $P_{\mathrm{A_{\rm 1}}}/P_{\mathrm{D}}=0.74623$. This period ratio is close, but significantly smaller than in High-amplitude $\delta$ Scuti (hereafter referred to as HADS, yellow pentagons in top Fig.~\ref{fig:petersen}). The second periodicity $P_{\mathrm{A_{\rm 2}}}=0.215331(2)$\,days with period ratio $P_{\mathrm{A_{\rm 2}}}/P_{\mathrm{D}}=0.71586$ fits into our newly discovered group. 

The basic properties of newly found double periodic stars are summarized in Tab.~\ref{tab:tab-see-app2}. The columns contain mean brightness $I$, period of the dominant, $P_{\mathrm{D}}$, and additional, $P_{\mathrm{A}}$, periodicity together with their period ratio, $P_{\mathrm{A}}/P_{\mathrm{D}}$. The following columns contain amplitudes of the dominant variability, $A_{\mathrm{D}}$, and the amplitude ratio of both periodicities, $A_{\mathrm{A}}/A_{\mathrm{D}}$. The last column, `remarks', contains notes on individual objects. The stars with unresolved signal near $P_{\mathrm{D}}$ (possible period change or long period Blazhko effect) are marked with `a'. Ten stars from our sample exhibit signatures of the Blazhko effect (quasi-periodic modulation of the light curve, for review, see \cite{Szab2014}). These stars are marked as `bl$_{\mathrm{D/A}}$', with regards whether the dominant and/or additional mode exhibits the Blazhko modulation. Stars that showed no combination frequencies were marked with `c'. The letter `d' marks stars with non-stationary additional periodicity. In a few stars, we have detected harmonic of the additional variability and mark them with `e'. The note about additional periodicity in \textit{remarks} file in the OGLE ftp archive is marked by `f'. 
In the top panel of Fig.~\ref{fig:ADPD} we show a distribution of amplitude ratios for our stars. In average the amplitude ratio vary in the $(5 - 90\,\%)$ range with most typical value 25 per cent. Stars from our group, previously classified as RRc stars, have higher amplitude ratio in comparison with the rest of our group of double-periodic variables. This is most likely caused by a lower amplitude of the dominant periodicity in stars identified as first-overtone variables. In the bottom panel of this figure, we show period-amplitude dependence for the dominant variability. We explicitly divided our stars into two groups based on OGLE classification. It is clearly visible that we have two groups, stars identified as RRab pulsators in one and six stars classified as RRc and RRab variables in the second.

\begin{table*}
\centering
\caption{Basic properties of discussed double-periodic stars. Subsequent columns provide star's id, mean $I$-band brightness, pulsation period of the dominant mode, the period of the additional mode, period ratio, amplitude of the dominant mode, amplitude ratio, and remarks. All stars were identified by OGLE as RRab stars, except the last four.}
\label{tab:tab-see-app2}
\begin{tabular}{llllllll}
\hline \hline
Name                 & \multicolumn{1}{c|}{$I$\,[mag]} & \multicolumn{1}{c|}{$P_{\mathrm{D}}$\,[days]} & \multicolumn{1}{c|}{$P_{\mathrm{A}}$\,[days]} & \multicolumn{1}{c|}{$P_{\mathrm{A}}/P_{\mathrm{D}}$} & \multicolumn{1}{c|}{$A_{\mathrm{D}}$\,[mag]} & \multicolumn{1}{c|}{$A_{\mathrm{A}}/A_{\mathrm{D}}$} & Remarks  \\ \hline 
OGLE-BLG-RRLYR-00595 & 16.770 & 0.28183 & 0.19971 & 0.70863 & 0.11414 & 0.287 &           \\
OGLE-BLG-RRLYR-00617 & 18.419 & 0.33811 & 0.24171 & 0.71491 & 0.16522 & 0.412 &           \\
OGLE-BLG-RRLYR-01401 & 16.892 & 0.31304 & 0.21390 & 0.68330 & 0.21913 & 0.056 &          \\
OGLE-BLG-RRLYR-01529 & 16.926 & 0.36332 & 0.25456 & 0.70064 & 0.17671 & 0.235 & bl$_{\mathrm{D}}$, f       \\
OGLE-BLG-RRLYR-02147 & 16.716 & 0.35221 & 0.25067 & 0.71171 & 0.17831 & 0.197 & a, f        \\
OGLE-BLG-RRLYR-02217 & 16.387 & 0.36557 & 0.25700 & 0.70301 & 0.11729 & 0.174 & f         \\
OGLE-BLG-RRLYR-02943 & 17.448 & 0.33669 & 0.22978 & 0.68247 & 0.18827 & 0.049 & a        \\
OGLE-BLG-RRLYR-02990 & 17.814 & 0.35305 & 0.24763 & 0.70139 & 0.15975 & 0.205 &         \\
OGLE-BLG-RRLYR-03090 & 17.197 & 0.32432 & 0.23087 & 0.71185 & 0.16619 & 0.184 & bl$_{\mathrm{A}}$       \\
OGLE-BLG-RRLYR-05851 & 18.105 & 0.35469 & 0.24563 & 0.69253 & 0.19680 & 0.108 &         \\
OGLE-BLG-RRLYR-07688 & 17.596 & 0.31544 & 0.21815 & 0.69159 & 0.19433 & 0.242 &         \\
OGLE-BLG-RRLYR-08621 & 16.242 & 0.33001 & 0.22940 & 0.69513 & 0.19538 & 0.124 &         \\
OGLE-BLG-RRLYR-09117 & 16.607 & 0.30968 & 0.21910 & 0.70750 & 0.18466 & 0.212 & bl$_{\mathrm{D}}$, bl$_{\mathrm{A}}$, e, f    \\
OGLE-BLG-RRLYR-12880 & 16.679 & 0.34004 & 0.23906 & 0.70302 & 0.21009 & 0.129 & bl$_{\mathrm{D}}$, d, e, f \\
OGLE-BLG-RRLYR-13745 & 16.391 & 0.35677 & 0.24762 & 0.69404 & 0.17592 & 0.213 & bl$_{\mathrm{A}}$       \\
OGLE-BLG-RRLYR-14063 & 16.239 & 0.37532 & 0.26343 & 0.70188 & 0.15565 & 0.155 & e, f        \\
OGLE-BLG-RRLYR-14135 & 17.571 & 0.30080 & 0.22447 & 0.74623 & 0.17557 & 0.173 &          \\
                     & 17.571 & 0.30080 & 0.21533 & 0.71586 & 0.17557 & 0.086 &         \\
OGLE-BLG-RRLYR-14481 & 16.271 & 0.40971 & 0.28202 & 0.68835 & 0.17284 & 0.255 & a        \\
OGLE-BLG-RRLYR-15657 & 15.993 & 0.36659 & 0.25402 & 0.69294 & 0.18889 & 0.289 &         \\
OGLE-BLG-RRLYR-15842 & 14.995 & 0.36544 & 0.25341 & 0.69345 & 0.17214 & 0.042 & a        \\
OGLE-BLG-RRLYR-16999 & 16.216 & 0.32582 & 0.23142 & 0.71026 & 0.17399 & 0.227 & bl$_{\mathrm{A}}$       \\
OGLE-BLG-RRLYR-18798 & 18.138 & 0.36868 & 0.26039 & 0.70628 & 0.18446 & 0.316 &          \\
OGLE-BLG-RRLYR-19058 & 17.703 & 0.30608 & 0.21404 & 0.69929 & 0.17930 & 0.195 &         \\
OGLE-BLG-RRLYR-19121 & 17.455 & 0.35760 & 0.24818 & 0.69402 & 0.15869 & 0.209 & bl$_{\mathrm{D}}$, f       \\
OGLE-BLG-RRLYR-20645 & 17.743 & 0.35648 & 0.24628 & 0.69088 & 0.17467 & 0.198 &          \\
OGLE-BLG-RRLYR-21400 & 17.922 & 0.35273 & 0.25389 & 0.71979 & 0.15273 & 0.354 & f         \\
OGLE-BLG-RRLYR-22305 & 18.465 & 0.34065 & 0.23816 & 0.69914 & 0.20571 & 0.240 & bl$_{\mathrm{A}}$       \\
OGLE-BLG-RRLYR-24360 & 17.920 & 0.32650 & 0.23003 & 0.70452 & 0.17587 & 0.207 &          \\
OGLE-BLG-RRLYR-24500 & 18.329 & 0.33858 & 0.24043 & 0.71011 & 0.17638 & 0.213 & c        \\
OGLE-BLG-RRLYR-27203 & 17.687 & 0.32119 & 0.22332 & 0.69527 & 0.16233 & 0.258 & bl$_{\mathrm{D}}$       \\
OGLE-BLG-RRLYR-28160 & 17.055 & 0.32689 & 0.23295 & 0.71261 & 0.16598 & 0.256 & f         \\
OGLE-BLG-RRLYR-29286 & 18.138 & 0.38828 & 0.27080 & 0.69744 & 0.17443 & 0.176 & c        \\
OGLE-BLG-RRLYR-30418 & 17.993 & 0.36410 & 0.25938 & 0.71238 & 0.18504 & 0.124 &          \\
OGLE-BLG-RRLYR-30470 & 17.244 & 0.38117 & 0.26376 & 0.69198 & 0.19405 & 0.131 &         \\
OGLE-BLG-RRLYR-30974 & 17.637 & 0.28973 & 0.20427 & 0.70505 & 0.20601 & 0.202 &          \\
OGLE-BLG-RRLYR-32721 & 17.046 & 0.32617 & 0.23016 & 0.70565 & 0.18223 & 0.210 & bl$_{\mathrm{A}}$       \\
OGLE-BLG-RRLYR-32923 & 18.246 & 0.34929 & 0.24503 & 0.70150 & 0.16659 & 0.274 & c        \\
OGLE-BLG-RRLYR-35838 & 16.065 & 0.31627 & 0.21852 & 0.69092 & 0.20416 & 0.235 & f         \\ \hline 
OGLE-BLG-RRLYR-19847 & 18.264 & 0.31772 & 0.22438 & 0.70624 & 0.11475 & 0.502 & c, f        \\
OGLE-BLG-RRLYR-30908 & 18.610 & 0.28746 & 0.20314 & 0.70668 & 0.10017 & 0.396 & f         \\
OGLE-BLG-RRLYR-31754 & 18.009 & 0.31732 & 0.22598 & 0.71216 & 0.08845 & 0.906 & e, f        \\
OGLE-BLG-RRLYR-34006 & 17.973 & 0.31207 & 0.21946 & 0.70323 & 0.08899 & 0.511 & f  \\    \hline 
\multicolumn{8}{l}{a - unresolved signal at $P_{\mathrm{D}}$; bl$_{\mathrm{D}}$ - peak(s) of the Blazhko effect in the vicinity of dominant mode; bl$_{\mathrm{A}}$ - peak(s) for the Blazhko effect } \\
\multicolumn{8}{l}{in the vicinity of additional mode; c - no combination frequency; d - non-stationary $P_{\mathrm{A}}$; e - harmonics of $P_{\mathrm{A}}$; f - listed in} \\
\multicolumn{8}{l}{ \textit{remarks} file in OGLE archive} \\
\hline \hline 
\end{tabular}
\end{table*} 

\begin{figure}
\centering
\resizebox{\hsize}{!}{\includegraphics{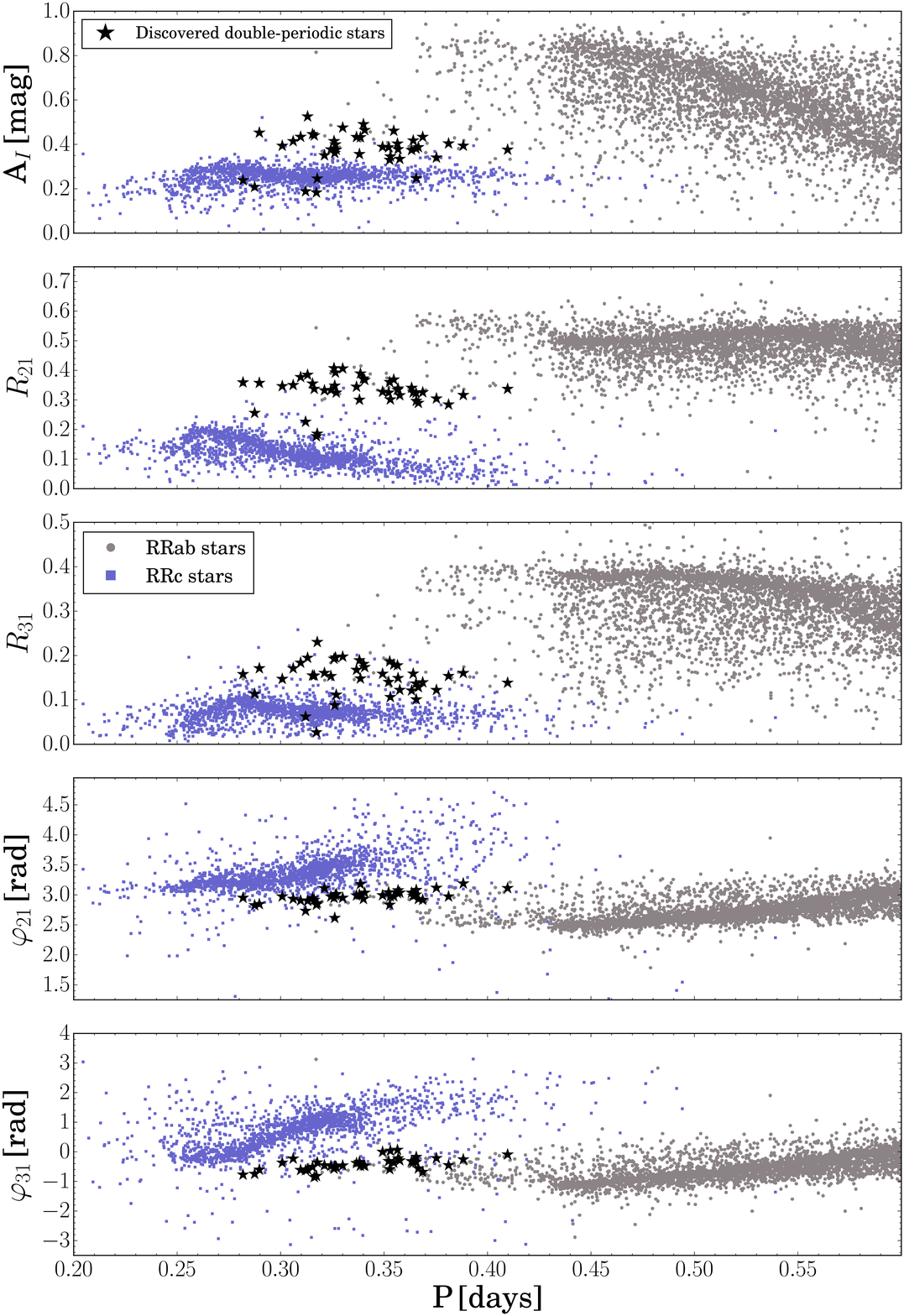}}
\caption{The Fourier decomposition parameters for the newly discovered double-periodic stars (black stars) with additional periodicity filtered out. For comparison, the Fourier coefficients for RRab stars (gray dots) and RRc stars (blue dots) are included.}
\label{fig:depen-koef}
\end{figure}

\begin{figure*}
\centering
\resizebox{\hsize}{!}{\includegraphics{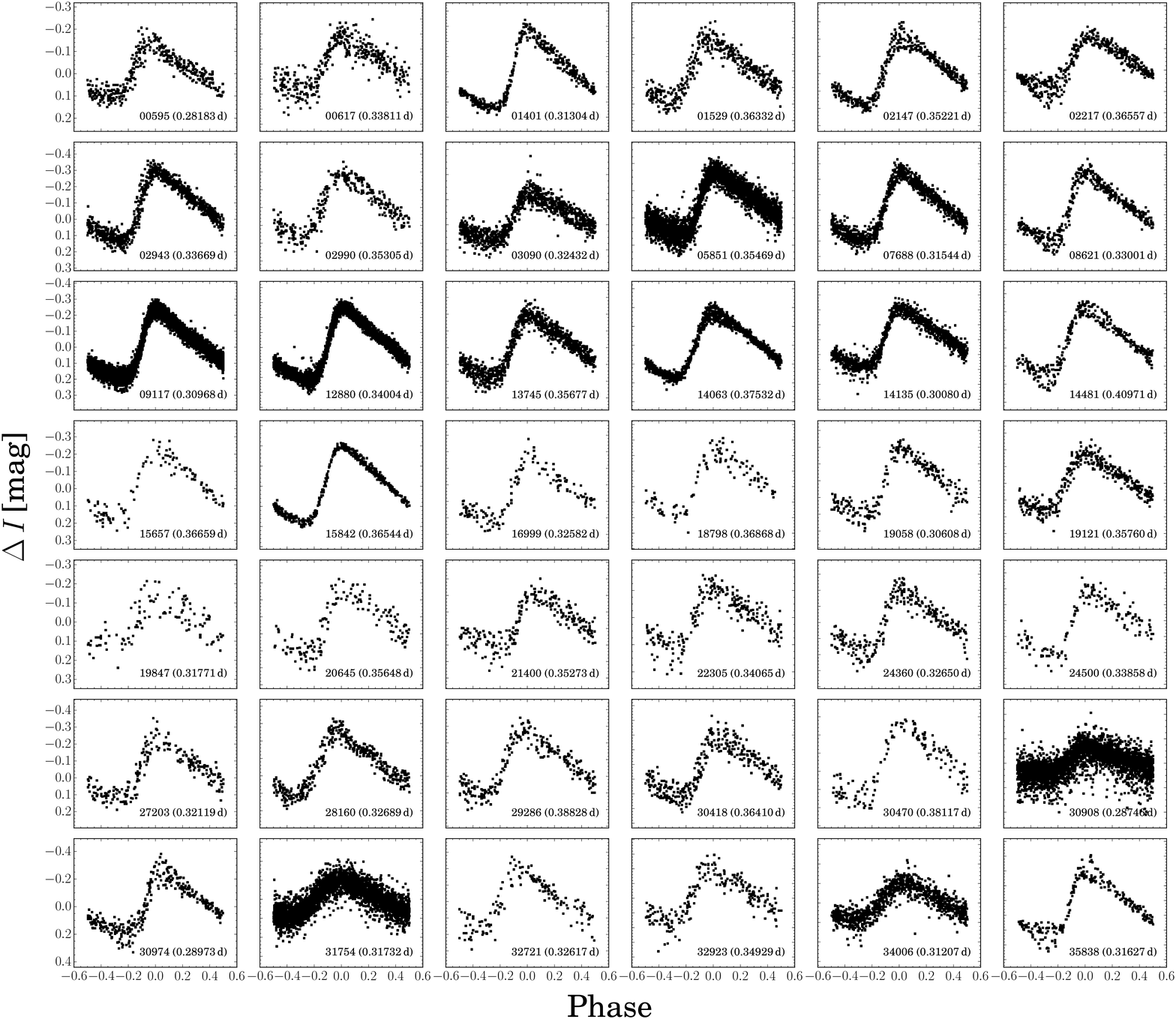}}
\caption{Light curves of the sample stars, phased with the dominant periodicity with the additional variability filtered out.}
\label{fig:phaseCurves}
\end{figure*}

Shapes of the light curves can be quantitatively described using the Fourier decomposition parameters \citep{Simon1981}. In Fig.~\ref{fig:depen-koef}, we compare the differences in our stars, RRc and RRab stars from the OGLE-IV Galactic bulge catalog \cite{ogleIV_rrl_blg}. We see, that the vast majority of our stars with extra periodicity fall in the region between RRc and RRab stars in the period-amplitude diagram. Four of our stars, identified by the OGLE team as RRc, fall in the middle of RRc group, which likely motivated their classification. However, their Fourier parameters are different, as we discuss in the following. Stars from the newly discovered group have, on average, shorter periods than RRab stars and larger amplitudes than RRc stars. For the Fourier amplitude ratios, $R_{\mathrm{21}}$ and $R_{\mathrm{31}}$, our stars populate the region between RRc and RRab stars. The Fourier phases $\varphi_{\mathrm{21}}$ and $\varphi_{\mathrm{31}}$ for our stars follow linear progression separating RRab and RRc stars. In addition, phase coefficients are among the lowest in the first-overtone stars. 

To explore the true nature of the dominant variability among these stars, we prewhitened the data with all components of the frequency spectra except $k\nu_{\rm D}$. The observations with removed additional periodicity phased with the dominant period, are displayed in Fig.~\ref{fig:phaseCurves}. The corresponding legend shows OGLE id and pulsation period of the dominant variability. All of the displayed light curves resemble triangular shape characteristic for the fundamental mode pulsators. 


Based on plots \ref{fig:depen-koef} and \ref{fig:phaseCurves} we suspect that the cause of the dominant periodicity is the same and thus fundamental pulsation mode seems to be the most probable for all stars from our sample (see Section \ref{sec:discussion} for the discussion).

Several stars from our double-periodic group seem to manifest the Blazhko modulation (listed in Tab.~\ref{tab:BlazhkoINRRf}). The columns contain OGLE identification, specification which periodicity is modulated ($P_{\mathrm{A}}$ or $P_{\mathrm{D}}$), period of the Blazhko modulation $P_{\mathrm{B}}$. Furthermore, we provide the amplitudes of the lower- and higher- frequency side peaks, $A_{-}$ and $A_{+}$, respectively, in the case they were detected in the frequency spectrum. In addition, we enclose a list of frequencies of detected modulation side peaks. The periods of the Blazhko modulation range from 7\,days up to almost 500\,days. In seven stars, triplet structures are detected in the frequency spectrum. OGLE-BLG-RRLYR-12880 and OGLE-BLG-RRLYR-09117 exhibit two Blazhko modulations. The OGLE-BLG-RRLYR-09117 displays modulation with the same period of the dominant and additional periodicity. Among the majority of the stars we see a preference for symmetry of the side peaks, or the $A_{+}$ component is higher. Only in OGLE-BLG-RRLYR-03090 solely the $A_{-}$ component is detected. It is worth noting that all stars listed in Tab.\,\ref{tab:BlazhkoINRRf} were previously classified as the fundamental mode pulsators.

\begin{table*}
\centering
\caption{List of double-periodic stars exhibiting the Blazhko modulation. Columns contain information about modulated periodicity, whether doublet or triplet structure is present and complete list of identified side peaks.}
\label{tab:BlazhkoINRRf}
\begin{tabular}{lclccl}
\hline \hline 
\multicolumn{1}{c}{OGLE-id} & mod. periodicity & $P_{B}$\,[days] & $A_{-}$\,[mag] & $A_{+}$\,[mag] & list of additional modulation side-peaks                                                         \\ \hline 
OGLE-BLG-RRLYR-01529        & $P_{\mathrm{D}}$   & 21.80(2)            & 0.007(1)           & 0.013(1)           & $2\nu_{\mathrm{D}}+\nu_{\mathrm{B}}$, $3\nu_{\mathrm{D}}+\nu_{\mathrm{B}}$ \\
OGLE-BLG-RRLYR-03090        & $P_{\mathrm{A}}$   & 121.8(5)            & 0.021(2)           & -                  & $\nu_{\mathrm{D}}+\nu_{\mathrm{A}} - \nu_{\mathrm{B}}$, $2\nu_{\mathrm{D}}+\nu_{\mathrm{A}}-\nu_{\mathrm{B}}$  \\
OGLE-BLG-RRLYR-09117        & $P_{\mathrm{D}}$   & 252.8(7)            & 0.0016(3)          & 0.0019(3)          & $\nu_{\mathrm{D}}+\nu_{\mathrm{B}}$, $\nu_{\mathrm{D}}-\nu_{\mathrm{B}}$, $2\nu_{\mathrm{D}}+\nu_{\mathrm{B}}$ \\
                            & $P_{\mathrm{A}}$   & 252.8(7)            & 0.0017(3)          & 0.0041(3)          & $\nu_{\mathrm{D}}+\nu_{\mathrm{A}} + \nu_{\mathrm{B}}$, $2\nu_{\mathrm{D}}+\nu_{\mathrm{A}}+\nu_{\mathrm{B}}$, $\nu_{\mathrm{D}}+\nu_{\mathrm{A}}-\nu_{\mathrm{B}}$, \\
                            & & & & &  $2\nu_{\mathrm{D}}+\nu_{\mathrm{A}}-\nu_{\mathrm{B}}$, $3\nu_{\mathrm{D}}+\nu_{\mathrm{A}}+\nu_{\mathrm{B}}$, $\nu_{\mathrm{D}}+2\nu_{\mathrm{A}}+\nu_{\mathrm{B}}$, \\
                            & & & & &  $\nu_{\mathrm{A}}-\nu_{\mathrm{D}}+\nu_{\mathrm{B}}$                      \\
OGLE-BLG-RRLYR-12880        & $P_{\mathrm{D}}$   & 70.85(8)            & 0.017(4)     &   0.019(7)              &                                                              $\nu_{\mathrm{D}}-\nu_{\mathrm{B_{1}}}$, $\nu_{\mathrm{D}}+\nu_{\mathrm{B_{1}}}$, $2\nu_{\mathrm{D}}-\nu_{\mathrm{B_{1}}}$, $2\nu_{\mathrm{D}}+\nu_{\mathrm{B_{1}}}$, \\
 & & & & & $3\nu_{\mathrm{D}}-\nu_{\mathrm{B_{1}}}$,   
$3\nu_{\mathrm{D}}+\nu_{\mathrm{B_{1}}}$, $4\nu_{\mathrm{D}}-\nu_{\mathrm{B_{1}}}$, $4\nu_{\mathrm{D}}+\nu_{\mathrm{B_{1}}}$,\\
 & & & & &  $5\nu_{\mathrm{D}}-\nu_{\mathrm{B_{1}}}$, $6\nu_{\mathrm{D}}-\nu_{\mathrm{B_{1}}}$, 
$5\nu_{\mathrm{D}}+\nu_{\mathrm{B_{1}}}$ \\
                            & $P_{\mathrm{D}}$   & 75.24(9)                    &   0.0056(4)                 &                   0.0055(4) &  $\nu_{\mathrm{D}}-\nu_{\mathrm{B_{2}}}$, $\nu_{\mathrm{D}}+\nu_{\mathrm{B_{2}}}$, $2\nu_{\mathrm{D}}-\nu_{\mathrm{B_{2}}}$, $2\nu_{\mathrm{D}}+\nu_{\mathrm{B_{2}}}$, \\
 & & & & & $3\nu_{\mathrm{D}}+\nu_{\mathrm{B_{2}}}$,  $3\nu_{\mathrm{D}}-\nu_{\mathrm{B_{2}}}$
                                                                                     \\
OGLE-BLG-RRLYR-13745        & $P_{\mathrm{A}}$   & 334.9(8)            & 0.020(1)           & 0.028(1)           & $\nu_{\mathrm{D}}+\nu_{\mathrm{A}}+\nu_{\mathrm{B}}$, $\nu_{\mathrm{D}}+\nu_{\mathrm{A}}-\nu_{\mathrm{B}}$, $2\nu_{\mathrm{D}}+\nu_{\mathrm{A}} + \nu_{\mathrm{B}}$,\\
                            & & & & &  $\nu_{\mathrm{A}} - \nu_{\mathrm{D}}+\nu_{\mathrm{B}}$,  $2\nu_{\mathrm{D}}+nu_{\mathrm{A}}-\nu_{\mathrm{B}}$, $3\nu_{\mathrm{D}}+\nu_{\mathrm{A}}-\nu_{\mathrm{B}}$, \\
                            & & & & &$\nu_{\mathrm{A}}-\nu_{\mathrm{D}}-\nu_{\mathrm{B}}$, $3\nu_{\mathrm{D}}+\nu_{\mathrm{A}}+\nu_{\mathrm{B}}$,    $4\nu_{\mathrm{D}}+\nu_{\mathrm{A}}-\nu_{\mathrm{B}}$,\\
                            & & & & & $3\nu_{\mathrm{D}}+2\nu_{\mathrm{A}}-\nu_{\mathrm{B}}$\\
OGLE-BLG-RRLYR-16999        & $P_{\mathrm{A}}$   & 507(11)             & -                  & 0.035(3)           &                                                                                    \\
OGLE-BLG-RRLYR-19121        & $P_{\mathrm{D}}$   & 482(5)              & 0.022(2)           & 0.018(2)           & $\nu_{\mathrm{D}}+\nu_{\mathrm{A}} - \nu_{\mathrm{B}}$, $\nu_{\mathrm{D}}+\nu_{\mathrm{A}}+\nu_{\mathrm{B}}$, $2\nu_{\mathrm{D}}+\nu_{\mathrm{A}}-\nu_{\mathrm{B}}$,\\
                            & & & & & $2\nu_{\mathrm{D}}+\nu_{\mathrm{A}}+\nu_{\mathrm{B}}$                                             \\
OGLE-BLG-RRLYR-22305        & $P_{\mathrm{A}}$   & 7.947(4)            & -                  & 0.030(6)           & $\nu_{\mathrm{D}}+\nu_{\mathrm{A}}+\nu_{\mathrm{B}}$ \\
OGLE-BLG-RRLYR-27203        & $P_{\mathrm{D}}$   & 37.09(4)            & 0.022(3)           & 0.019(3)           & $2\nu_{\mathrm{D}}+\nu_{\mathrm{B}}$ \\
OGLE-BLG-RRLYR-32721        & $P_{\mathrm{A}}$   & 495(16)             & -                  & 0.026(3)           & \\ \hline \hline 
\end{tabular}
\end{table*}

To summarize, we have discovered a new and distinct class of double periodic pulsators with the following characteristics:
\begin{itemize}
\item{Period of the dominant variability $0.28-0.41$\,days.}
\item{The additional periodicity is shorter.}
\item{The period ratios are in the range of $0.68 - 0.72$.}
\item{Typical amplitude ratio is 25\,\%.}
\item{Triangular shape of light curves suggests dominant fundamental mode for all stars from our sample.}
\item{Values of $R_{\mathrm{21}}$ and $R_{\mathrm{31}}$ lie between typical values for RRc and RRab stars.}
\item{The $\varphi_{\mathrm{21}}$ and $\varphi_{\mathrm{31}}$ coefficients form a linear progression between RRc and RRab stars.}
\item{Ten stars from our sample exhibit the Blazhko modulation, and either the dominant, additional or both periodicities are modulated.}
\end{itemize}

\subsection{Remarks on the most well-observed stars} \label{subsec:Remarks} 

This section contains notes on a few of the most well-observed stars from the OGLE photometry that exhibit extra periodicity. For each star, we present periods for the dominant and the additional variability together with their period ratio. In addition, we provide a list of combination frequencies that appeared in the frequency spectrum and other possible peculiarities of an individual object.

\begin{figure}
\centering
\resizebox{\hsize}{!}{\includegraphics{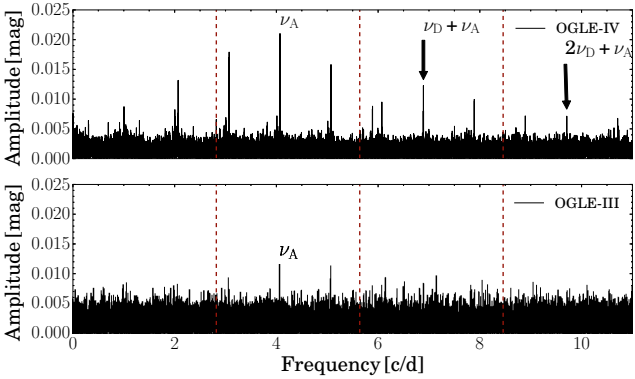}}
\caption{The top panel shows the frequency spectrum of the OGLE-IV data for OGLE-BLG-RRLYR-05851 after prewhitening with dominant frequency and its harmonics. Additional variability and combination frequencies of both periodicities are marked. Other significant peaks are their daily aliases. The bottom figure shows the frequency spectrum of the same star but for the OGLE-III data. Dashed red lines mark frequencies of the main pulsation frequency and its harmonics.}
\label{fig:5851-frspec}
\end{figure}

\textbf{OGLE-BLG-RRLYR-05851}. The pulsation period is close to the average within our group, $P_{\mathrm{D}}=0.35468824(22)$\,days, in addition we found $P_{\mathrm{A}}=0.2456316(12)$\,days. The period ratio $P_{\mathrm{A}}$/$P_{\mathrm{D}}=0.69253$ fall into the lower boundary of our group. This star has one of the lowest amplitudes for the additional periodicity (only 10 percent of the amplitude of the fundamental mode). In the frequency spectrum (Fig.~\ref{fig:5851-frspec}), we found five harmonics for the fundamental mode and two combination frequencies $\nu_{\mathrm{D}}+\nu_{\mathrm{A}}$ and $2\nu_{\mathrm{D}}+\nu_{\mathrm{A}}$. In Fig.~\ref{fig:5851-frspec}  we notice an obvious difference in the frequency spectrum for OGLE-III and OGLE-IV. We did not detect any combination frequencies in OGLE-III data. This could be caused by lower amplitude of the additional periodicity in the OGLE-III data.

\textbf{OGLE-BLG-RRLYR-09117}. The most often observed star in our sample, with the fifth shortest pulsation period $P_{\mathrm{D}}=0.30967728(3)$\,days. The additional variability has a period $P_{\mathrm{A}}=0.21909756(19)$\,days and period ratio $P_{\mathrm{A}}/P_{\mathrm{D}}=0.70750$ is typical for our group. 

\begin{figure}
\centering
\resizebox{\hsize}{!}{\includegraphics{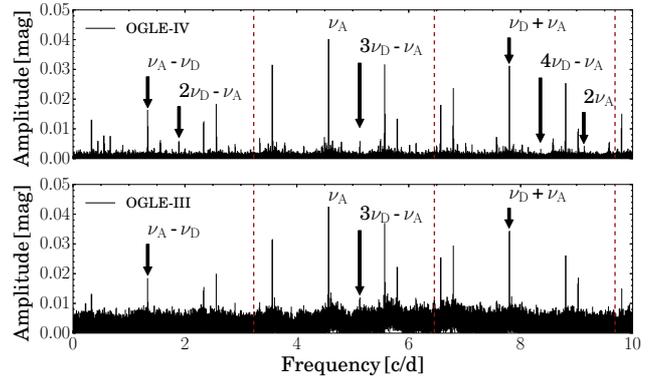}}
\caption{The same as Fig.\,$\ref{fig:5851-frspec}$ but for the OGLE-BLG-RRLYR-09117.}
\label{fig:9117-frspec}
\end{figure}

The frequency spectrum (Fig.~\ref{fig:9117-frspec}) contains eight harmonics for the fundamental mode, additional periodicity and its harmonic, and 14 combination frequencies in the following form $k\nu_{\mathrm{D}} + \nu_{\mathrm{A}}$ ($k=1,\ldots, 6$), $\nu_{\mathrm{A}}-\nu_{\mathrm{D}}$, $2\nu_{\mathrm{A}}-\nu_{\mathrm{D}}$, $k\nu_{\mathrm{D}}+2\nu_{\mathrm{A}}$ ($k=0,\ldots, 4$) and $k\nu_{\mathrm{D}}-\nu_{\mathrm{A}}$ ($k=2,\ldots, 5$). In addition, this star, as the only one from our sample, exhibits the Blazhko modulation with the same period in both periodicities, with period $P_{\mathrm{B}}=252.8(7)$\,days. In the frequency spectrum, we see a triplet and doublet structure in the vicinity of the additional periodicity and the fundamental mode, respectively (see Tab. \ref{tab:BlazhkoINRRf}). 

\begin{figure*}
\centering
\resizebox{\hsize}{!}{\includegraphics{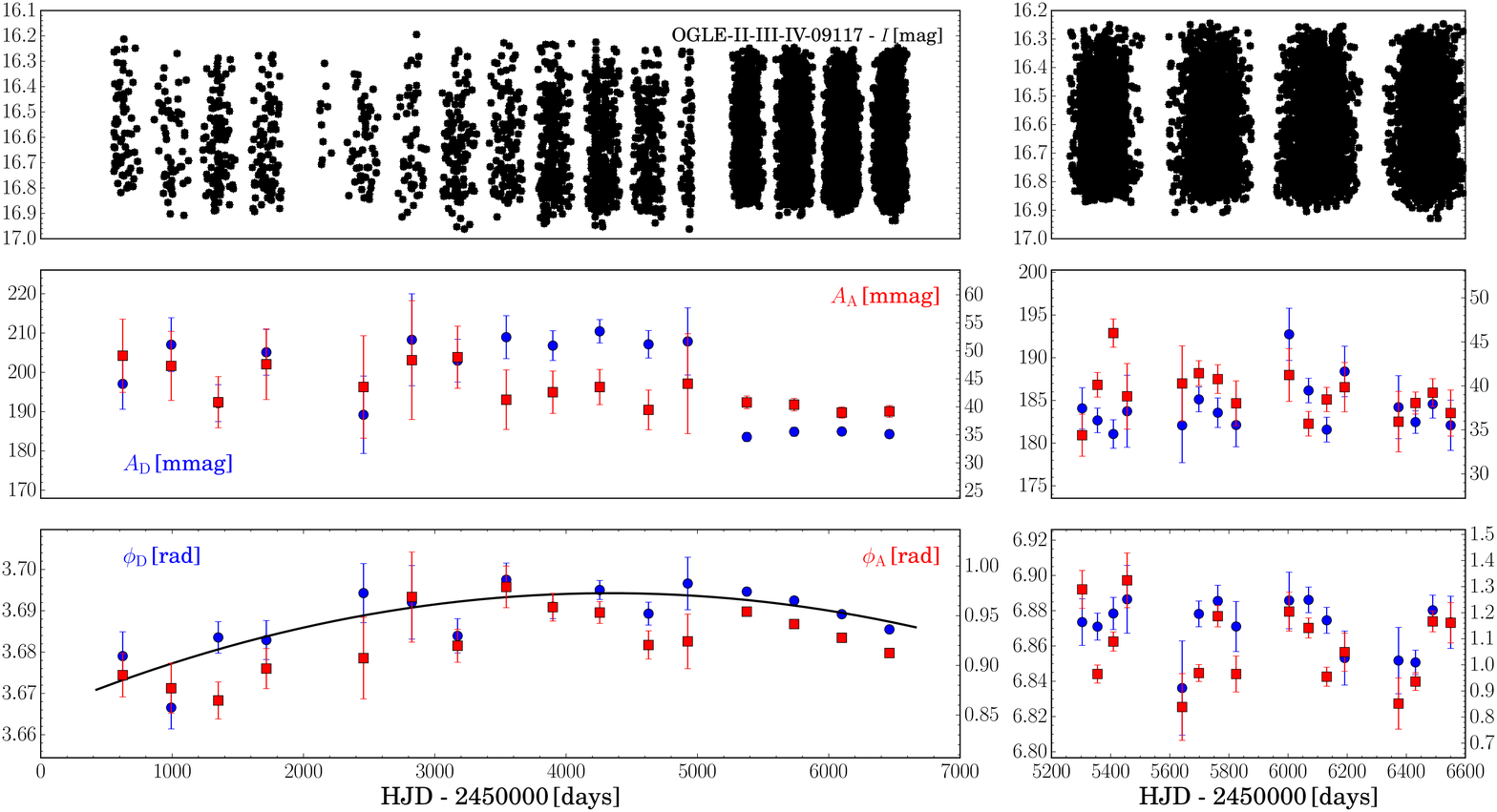}}
\caption{Time-dependent analysis of amplitude and phase changes of the two periodicities in OGLE-BLG-RRLYR-09117. The top panels show available photometric data in the OGLE catalogue. The middle and bottom panels display amplitude and phase changes for both variabilities. Squared symbols characterize the dominant periodicity. Its amplitude and phase correspond to the left-hand side vertical axes. Circular symbols characterize the additional variability. Its amplitudes and phase correspond to the right-hand side vertical axes. The left-hand panels show data analysed on season-to-season basis, while the right-hand panels show analysis of OGLE-IV data with resolution $\approx$ 50\,days.}
\label{fig:9117_tdfd}
\end{figure*}

The numerous and high-quality data allowed us to use the time-dependent Fourier analysis \citep{kbd87} using merged OGLE-II, OGLE-III and OGLE-IV data (see Fig.~\ref{fig:9117_tdfd}). The top panel shows available photometric data. Changes in amplitude and phase for a given periodicity are displayed in the middle and bottom panel. In the bottom plot, we clearly see phase changes in the dominant and additional periodicity. While for the later, the change is not so pronounced, the variation in the dominant mode can be modeled using a parabolic function, which gives us growing pulsation period with a rate of $1.5\,\cdot\,10^{-6}$\,days per 1\,000\,days.

\begin{figure}
\centering
\resizebox{\hsize}{!}{\includegraphics{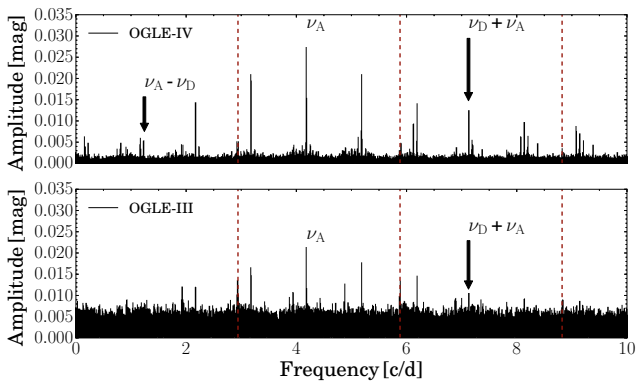}}
\caption{The same as Fig.~\ref{fig:5851-frspec} but for the OGLE-BLG-RRLYR-12880.}
\label{fig:12880-frspec}
\end{figure}

\textbf{OGLE-BLG-RRLYR-12880}. The star was identified as the fundamental mode pulsator, with pulsation period $P_{\mathrm{D}}=0.3400408(1)$\,days, furthermore we found extra periodicity $P_{\mathrm{A}}=0.2390565(5)$\,days. The period ratio $P_{\mathrm{A}}/P_{\mathrm{D}}=0.70302$ is similar to the other stars in our sample, but the amplitude ratio $A_{\mathrm{A}}/A_{\mathrm{D}}=0.1292$ is somewhat lower. The frequency spectrum (see Fig.~\ref{fig:12880-frspec}) contains six harmonics for the fundamental mode and one harmonic for the additional periodicity ($\nu_{\mathrm{A}}$ and $2\nu_{\mathrm{A}}$). In addition, we detected following combinations: $k\nu_{\mathrm{D}}+\nu_{\mathrm{A}}$ ($k=1,\ldots, 5$), $\nu_{\mathrm{A}}-\nu_{\mathrm{D}}$ and $2\nu_{\mathrm{A}}+2\nu_{\mathrm{D}}$. Two Blazhko modulations were detected in the vicinity of the fundamental mode, with periods $P_{\rm B1}=70.85(8)$\,days and $P_{\rm B2}=75.24(9)$\,days. In the frequency spectrum we see a residua at the $k\nu_{\mathrm{A}}$, which suggest changes in period of the additional variability. Similarly as in Fig.~\ref{fig:5851-frspec}, in Fig.~\ref{fig:12880-frspec} we see lower number of combination frequencies in the frequency spectrum computed from OGLE-III. This could be caused by lower amplitude of the additional periodicity.

\begin{figure}
\centering
\resizebox{\hsize}{!}{\includegraphics{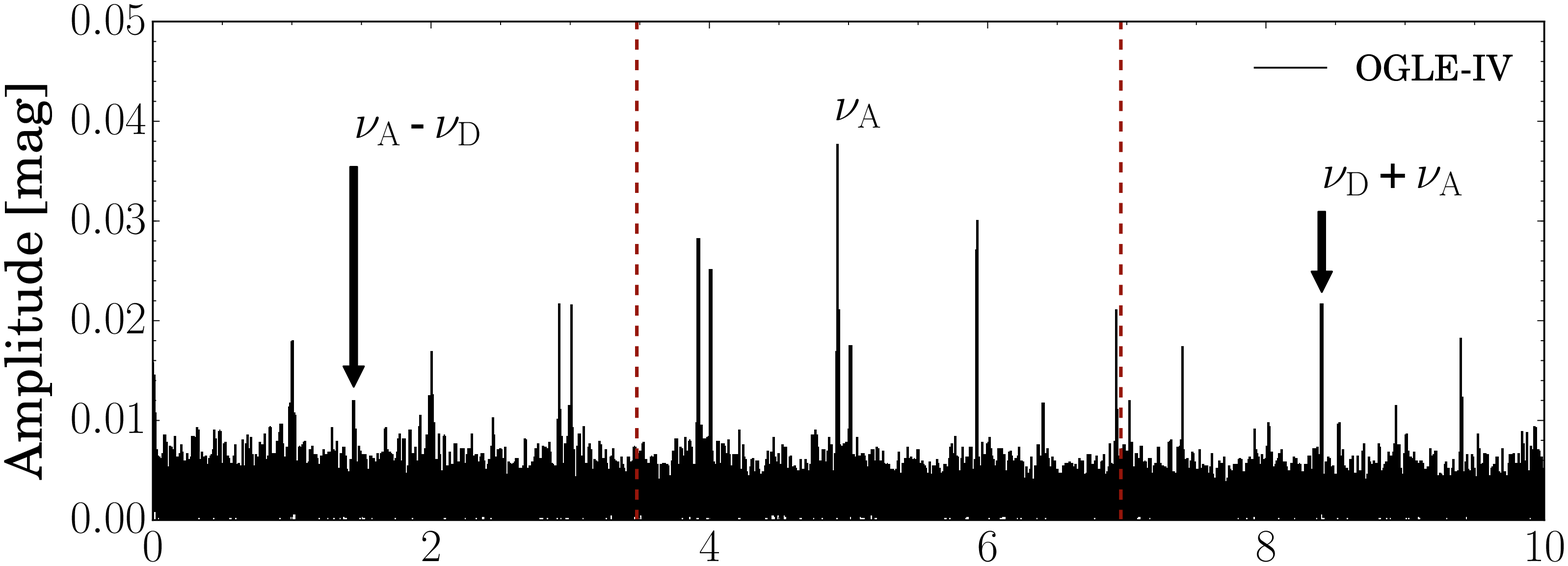}}
\resizebox{\hsize}{!}{\includegraphics{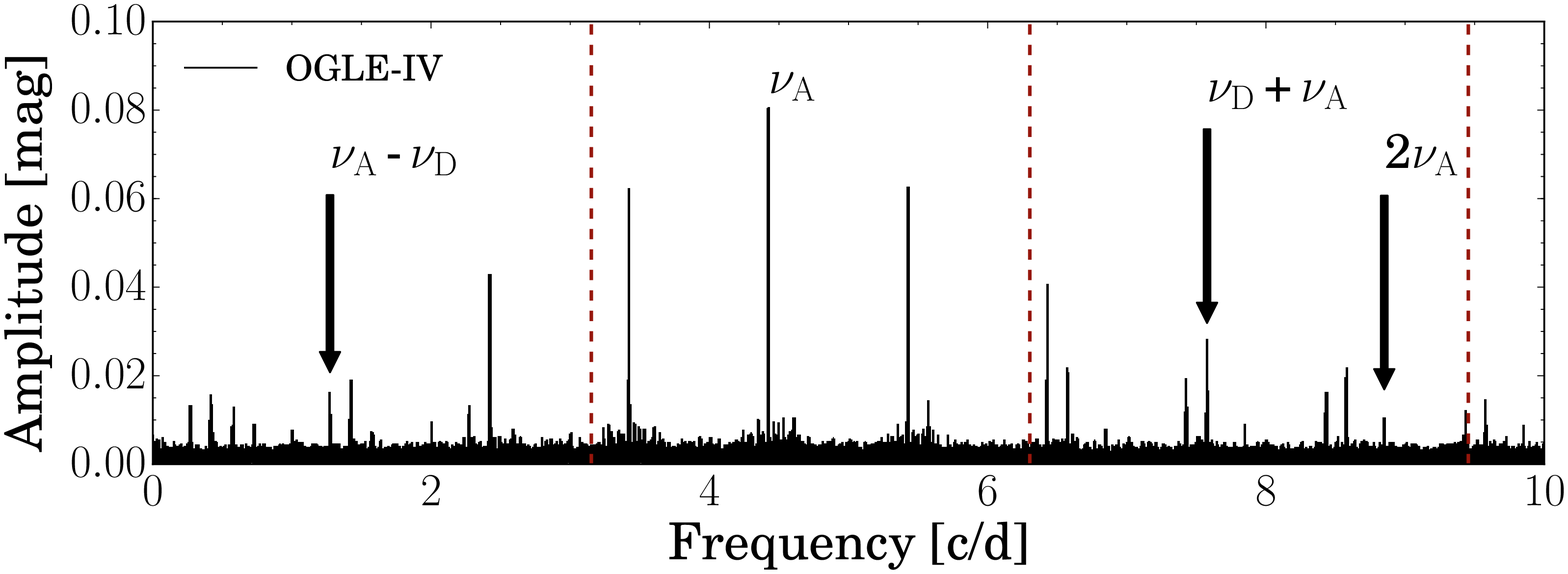}}
\caption{The frequency spectra for the OGLE-BLG-RRLYR-30908 (top panel) and for the OGLE-BLG-RRLYR-31754 (bottom panel) after prewhitening with dominant frequency and its harmonics. Additional periodicity and combination frequencies of both variabilities are marked. Other significant peaks are their daily aliases. Only OGLE-IV data were available. Dashed red lines mark frequencies of the main pulsation frequency and its harmonics.}
\label{fig:30908-31754-frspec}
\end{figure}

\textbf{OGLE-BLG-RRLYR-30908}. One of the four stars classified as RRc variable in our sample with pulsation period $P_{\mathrm{D}}=0.28746404(55)$\,days, the second shortest period among discovered stars and additional periodicity with period of $P_{\mathrm{A}}=0.20314449(70)$\,days. Thus the period ratio $P_{\mathrm{A}}/P_{\mathrm{D}}=0.70668$ fall in the middle of our group. The amplitude ratio $A_{\mathrm{A}}/A_{\mathrm{D}}=0.3958$ is the lowest among the stars identified by \cite{ogleIV_rrl_blg} as the first-overtone pulsators in our group. The frequency spectrum (see Fig.~\ref{fig:30908-31754-frspec} top plot) of this star contains three harmonics of the dominant periodicity. We found two combination frequencies of the additional variability with the dominant periodicity, $\nu_{\mathrm{D}}+\nu_{\mathrm{A}}$ and $\nu_{\mathrm{A}}-\nu_{\mathrm{D}}$.

\textbf{OGLE-BLG-RRLYR-31754}. Another star from our sample originally classified as RRc with dominant pulsation period $P_{\mathrm{D}}=0.31731570(42)$\,days and with additional periodicity $P_{\mathrm{A}}=0.22598012(25)$\,days (period ratio $P_{\mathrm{A}}/P_{\mathrm{D}}=0.71216$). The amplitude ratio $A_{\mathrm{A}}/A_{\mathrm{D}}=0.9058$ is the highest among stars in our group. The frequency spectrum displays two harmonics for the dominant periodicity and one for the additional variability (see Fig.~\ref{fig:30908-31754-frspec} bottom plot). Furthermore, frequency spectrum of this star contains following combinations: $\nu_{\mathrm{D}}+\nu_{\mathrm{A}}$, $2\nu_{\mathrm{D}}+\nu_{\mathrm{A}}$, $\nu_{\mathrm{D}}+2\nu_{\mathrm{A}}$, $\nu_{\mathrm{A}}-\nu_{\mathrm{D}}$.

Based on figures displaying frequency spectrum in subsection \ref{subsec:Remarks} and scarce appearance of the mark `a' in Tab.~\ref{tab:tab-see-app2}, it is worth mentioning that the dominant periodicity is stable in the vast majority of cases. Only in four instances, we detected residual power at the $k\nu_{\mathrm{D}}$ after the prewhitening, which suggests changes in the period and/or amplitude (mark `a' in Tab.~\ref{tab:tab-see-app2}).

\section{Discussion}\label{sec:discussion}

We have discovered a new group of double-periodic variables. Majority of these stars are identified in the OGLE catalog as RRab -- RR~Lyr stars pulsating in the radial fundamental mode. The characteristic, asymmetric light curve shape (Fig.~\ref{fig:phaseCurves}) leaves no doubt -- it must correspond to pulsation. The large amplitude and non-linear, triangular shape of the light curve immediately suggests the radial fundamental mode. A glimpse at the Fourier parameter plots (Fig.~\ref{fig:depen-koef}) points however, that this is not the light curve shape that is typical for RRab stars with such short pulsation periods. In fact, genuine RRab stars with periods shorter than $0.43$\thinspace days are scarce in the Galactic bulge and nearly not present for periods shorter than $0.35$\thinspace days. The light curves do not resemble the RRc light curves either. The Fourier parameters corresponding to the dominant variability lie {\it in between} the parameters typical for RRab and RRc stars.

\begin{figure}
\centering
\resizebox{\hsize}{!}{\includegraphics{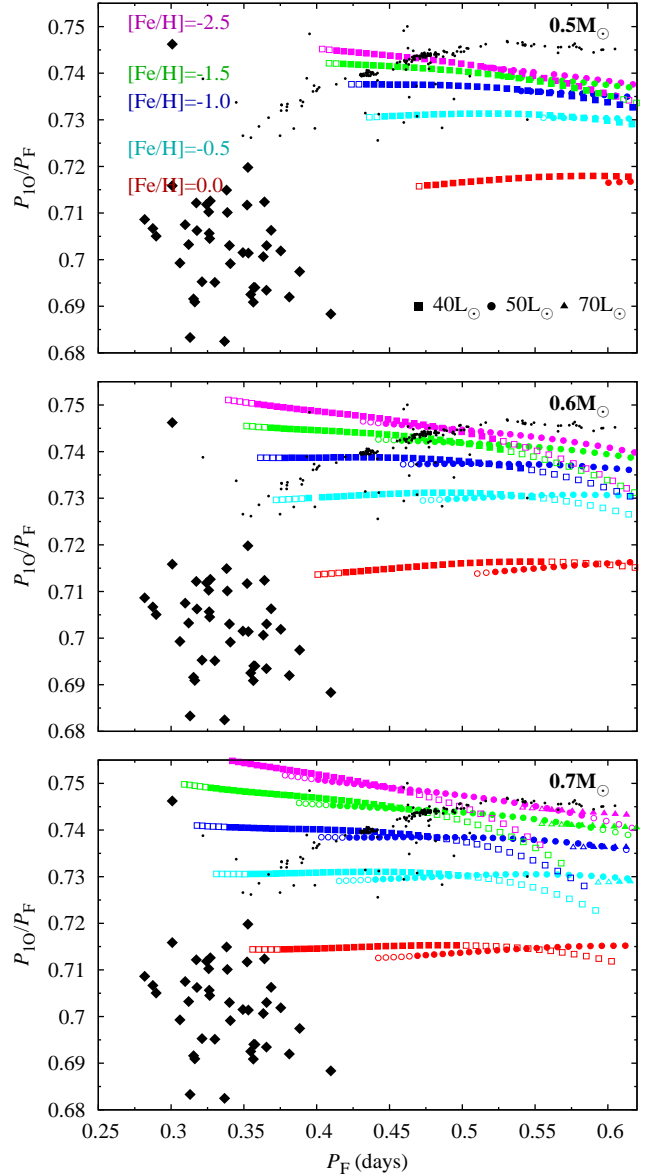}}
\caption{Modelled radial mode period ratios confronted with the new double-periodic variables in the Petersen diagram. RRd stars from the Galactic bulge are also plotted for reference (small dots). Three panels correspond to different model masses, as indicated in the top right corners. Model sequences are plotted with different symbols, depending on the models' luminosity and with different colors, depending on the models' metallicity, as indicated in the top panel. Filled symbols correspond to the models in which fundamental mode and first-overtone are simultaneously linearly unstable. Open symbols correspond to the models in which only one mode is linearly unstable.}
\label{fig:models}
\end{figure}

The secondary variability in the discussed stars, is always of shorter period and of lower amplitude as compared to the dominant variability. The corresponding light curve shape is a pure sine wave in nearly all the cases, i.e. we do not detect the harmonics of the additional variability. The secondary period is too short to be connected with rotation or with the binarity effects. In the Petersen diagram (Fig.~\ref{fig:petersen}), the newly identified group forms a rather large cloud that touches the RRd sequence at its extreme, short-period tail. We note in passing, that this tail is observed only in the Galactic bulge, and explained as due to high metallicity RRd variables \citep{ogleIII_rrl_blg}. Could the discussed stars be extreme cases of RRd variables, with the radial fundamental and radial first-overtone modes excited? Such possibility can be easily verified with linear pulsation calculations. The results are presented in Fig.~\ref{fig:models}. We used exactly the same set of models as in \cite{smolec2016} and the reader is referred to its sect.~4.1 for all technical details. The filled symbols mark the models in which both radial modes are simultaneously unstable -- a necessary condition for the non-resonant beat pulsation. Open symbols to the left, correspond to the models in which only the first-overtone is unstable and open symbols to the right, correspond to the models in which only the fundamental mode is unstable. Although the models cover a wide range of metallicities, masses and luminosities, it is clear that our group cannot correspond to RRd pulsation. Only for the highest (solar) metallicity, the highest mass ($0.7\MS$), and the lowest luminosity ($40\LS$), the models in which both radial modes are simultaneously unstable approach the part of the Petersen diagram in which the new, double-periodic variables reside. The extreme RRd scenario must be dropped. The models also indicate, that for the considered range of physical parameters, it is even hard to get the unstable fundamental mode at such short periods.

Recently, \cite{Soszysky2016MNRAS} identified a new group of anomalous RRd variables. These stars however, have longer fundamental mode periods and higher period ratios (still, typically lower than in the ``classical'' RRd stars), so their position in the Petersen diagram is easily reproduced assuming excitation of radial fundamental and first-overtone modes. In addition, Blazhko-like modulation is commonly detected in anomalous RRd stars. Variables discussed in this paper clearly do not belong to this group.

\begin{figure}
\centering
\resizebox{\hsize}{!}{\includegraphics{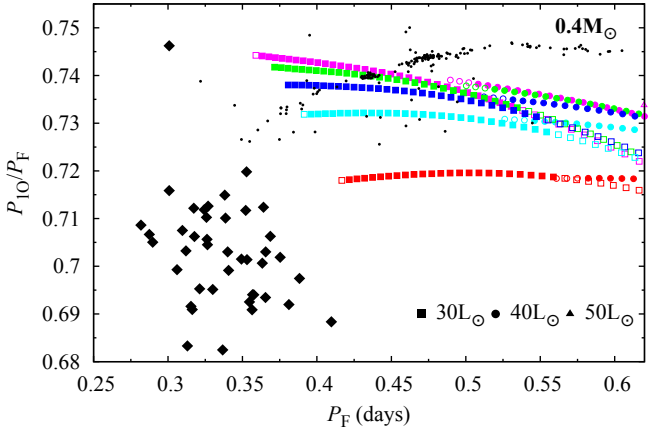}}
\caption{Similar to Fig.~\ref{fig:models}, but for BEP models of $0.4\MS$. The different colours correspond to different metallicities, as indicated in the top panel of Fig.~\ref{fig:models}.}
\label{fig:modelsBEP}
\end{figure}

The other possibility we now consider, is that the discussed stars are the so-called binary evolution pulsators (BEPs). The first, and so far the only member of this group, is OGLE-BLG-RRLYR-02792. The star was catalogued as RRab, member of the eclipsing binary system \citep{ogleIII_rrl_blg}. Its dynamical mass is only $0.26\MS$  however \citep{bepnat}, definitely too low for RR~Lyr variable. BEPs are stars that evolved in the binary system and, after experiencing the mass loss event, enter the classical instability strip. The non-linear model survey conducted by \cite{bep} shows that such objects may indeed show large-amplitude radial pulsation that may mimic RR~Lyr variability. In fact, models presented in \cite{bep} seem to support the hypothesis that the discussed double-periodic stars are BEPs. The expected fundamental mode light curves at the blue edge of the classical instability strip have simple triangle-like form \citep[fig.~11 in][]{bep}. The Fourier parameters of hot BEPs fall in between RRab and RRc stars (their fig.~13; beginning of the model sequences). Also, close to the blue edge, both low order radial modes may be simultaneously excited (their fig.~3) and beat pulsation is thus possible. The linear pulsation models disprove this scenario as well. We have computed a grid of models with masses in the $0.2-0.4\MS$ range, with luminosities in the $20-70\LS$ range and metallicity range the same as considered in Fig.~\ref{fig:models} \citep[other model parameters are the same as in][]{smolec2016}. We note that little is known about expected physical parameters of BEPs, as only one object was observed so far \citep[see however][]{karczmarek}. The range of the physical parameters given above, is motivated by the model study by \cite{bep}, and assures that model periods are not far from those observed in RR~Lyr stars. The models assume homogeneous envelope with chemical composition the same as in RR~Lyr-type models. For $0.2\MS$ and $0.3\MS$ models, periods of the unstable radial modes are significantly longer than observed in our stars. Period ratios for $0.4\MS$ models are plotted in Fig.~\ref{fig:modelsBEP}. The models have longer periods and higher period ratios, and are far from the region in which double-periodic stars reside. Clearly, the discussed stars are not fundamental and first-overtone double-mode radial pulsators. At least, not assuming their physical parameters are close to expected for RR~Lyr stars or BEPs (that mimic RR~Lyr stars). Other combinations of radial modes do not fit the discussed group as well.

We note that simple triangular light curve shape is characteristic for fundamental mode HADS stars. Fourier parameters of these stars are similar as well \citep[see fig. 8 and figs. 4-7 in][note that cosine convention is adopted in that paper]{poretti}. Periods of these stars are typically below $0.2$\thinspace days however, while in our stars they are longer than $0.28$\thinspace days. Period ratios in the radial multi-mode HADS stars are also vastly different, see e.g. fig.~10 in \cite{Pietrukowicz2013}.

Assuming that the dominant variability do correspond to the radial fundamental mode (large amplitude and nonlinear light curve shape make this hypothesis most plausible), the additional variability may correspond to non-radial pulsation mode. Low-degree modes may be excited in RR~Lyr models \citep{vdk98,dc99}, preferentially in the vicinity of the radial modes. Although the frequency spectrum of such modes is very dense, their growth rates do display a local maximum close to the position of the radial modes. In particular, for $\ell=1$ modes, the growth rates approach that of the radial first-overtone mode and their frequency is higher than radial first-overtone frequency [see e.g. fig.~1 in \cite{vdk98} or fig.~1 in \cite{wd16}]. This would lead to the period ratios a bit lower than observed for RRd variables, which is what we want. Fig.~\ref{fig:models} seems to disprove this scenario again. Non-radial modes of $\ell=1$ are unstable only if the first-overtone is unstable, which occurs at longer fundamental mode periods than observed in the discussed stars. Also, it would not be clear why the non-radial mode is excited instead of the neighbouring, strongly driven radial first-overtone.

At the moment we have no explanation for the nature of the discussed double-periodic variables.

\section{Conclusions}\label{sec:Conclusions}
We present the discovery of a new group of double-periodic stars in the OGLE Galactic bulge photometry. In 38 stars identified in the OGLE catalog as the fundamental mode RR Lyrae stars and in 4 stars identified as the first-overtone pulsators (RRc), we detect additional variability of lower amplitude and of a shorter period. The period ratios in the discovered stars ($0.68 - 0.72$) are lower than in the Galactic bulge RRd stars ($0.725-0.750$). 

Although our stars do not form a sharp sequence in the Petersen diagram, they occupy a well-defined region. Also, their light curves are very similar and different from the light curves of single-periodic RRab and RRc stars. This is best demonstrated with the help of Fourier decomposition parameters. The Fourier coefficients $R_{\mathrm{21}}$, $R_{\mathrm{31}}$, $\varphi_{\mathrm{21}}$ and $\varphi_{\mathrm{31}}$ for discovered stars differ from typical values for stars in RRab and RRc group. For the Fourier parameters $R_{\mathrm{21}}$ and $R_{\mathrm{31}}$, our group lies somewhere in the middle, between single periodic RRc and RRab stars. The newly discovered double-periodic stars have higher values of these coefficients in comparison with RRc stars. The similar difference applies for the phase coefficients $\varphi_{\mathrm{21}}$ and $\varphi_{\mathrm{31}}$ where our group creates some kind of a transition between the fundamental mode and first-overtone pulsators. In addition, stars in our group have the lowest values of these coefficients among RRc stars. Our findings suggest that the original OGLE classification as RRab or RRc is questionable in our stars. Ten stars from our sample show properties typical for Blazhko-type modulation (equidistant peaks in the frequency spectrum). In the majority of modulated stars, the modulation side peaks are symmetrical, or we observe that higher frequency side peak is of higher amplitude.

The additional periodicity is always of lower amplitude. Except for one case, amplitude ratios are below 0.6; the most typical value is 0.2 (see the top panel of Fig.~\ref{fig:ADPD}). Considering the amplitude of the dominant variability (see the lower panel of Fig.~\ref{fig:ADPD}), our stars divide into two groups: with larger amplitude ($A_{\rm D}>0.15$\,mag, all stars originally classified as RRab, except two) and with lower amplitude ($A_{\rm D}<0.12$\,mag, four stars originally classified as RRc and two as RRab). Origin of the apparent gap in the distribution of amplitudes of the dominant variability is unclear.

We find it very challenging to explain the nature of the discovered group. Although these stars were identified as RR~Lyrae, their properties are very different from single periodic RRab and RRc stars and from double-mode RRd stars. RR~Lyrae identification is thus tentative. The triangular shape of the light curve of the dominant periodicity strongly suggests that it corresponds to the pulsation in the radial fundamental mode. However, the light curve shape is clearly different from that observed in RRab stars. Period ratios are also significantly smaller than observed in RRd stars. With linear pulsation models we have checked, that location of these stars in the Petersen diagram cannot be reproduced assuming that two radial modes are excited and assuming that their physical parameters are in the range expected for RR~Lyrae stars or the so-called binary evolution pulsators (masses $0.2-0.7\MS$, luminosities in the $20-70\LS$ range, metallicities, [Fe/H] from 0 to -2.5). The scenario in which dominant variability corresponds to the radial fundamental mode and additional variability corresponds to non-radial mode also faces difficulty.

\section*{Acknowledgements}
We are grateful to Wojtek Dziembowski and Pawel Moskalik for fruitful discussions concerning the nature of described variables. RS and HN are supported by the National Science Center, Poland, grants no. DEC-2012/05/B/ST9/03932 and DEC-2015/17/B/ST9/03421. MS acknowledges the support of the postdoctoral fellowship programme of the Hungarian Academy of Sciences at the Konkoly Observatory as a host institution, and the financial support of the Hungarian NKFIH Grant K-115709. 





\begin{thebibliography}{99}

\bibitem[\protect\citeauthoryear{Dziembowski \& Cassisi}{1999}]{dc99} Dziembowski W.A., Cassisi S., 1999, Acta Astron., 49, 371
\bibitem[\protect\citeauthoryear{Dziembowski}{2016}]{wd16} Dziembowski W., 2016, Comm. Konkoly Obs., vol. 105, 23
\bibitem[\protect\citeauthoryear{Gruberbauer et al.}{2007}]{aqleo} Gruberbauer M., Kolenberg K., Rowe J. et al., 2007, MNRAS, 379, 1498
\bibitem[\protect\citeauthoryear{Herzberg \& Glogowski}{2014}]{Herzberg2014} Herzberg W., Glogowski K., 2014, IAUS, 301, 421
\bibitem[\protect\citeauthoryear{Jurcsik et al.}{2014}]{Jurcsik2014} Jurcsik J., Smitola P., Hajdu G., Nuspl J., 2014, ApJ, 797, L3 
\bibitem[\protect\citeauthoryear{Karczmarek et al.}{2017}]{karczmarek} Karczmarek P., Wiktorowicz G., Ilkiewicz K., Smolec R., Stepie\'{n} K., Pietrzy\'{n}ski G., Gieren W., Belczynski K., 2017, MNRAS, submitted
\bibitem[\protect\citeauthoryear{Kov\'acs, Buchler \& Davis}{1987}]{kbd87}  Kov\'acs G., Buchler J.R., Davis C.G., 1987, ApJ, 319, 247
\bibitem[Molnar(2016)]{molnar2016} Molnar, L.\ 2016, Commmunications of the Konkoly Observatory Hungary, 105, 11
\bibitem[\protect\citeauthoryear{Moskalik et al.}{2015}]{pamsm15} Moskalik P., Smolec R., Kolenberg K. et al., 2015, MNRAS, 447, 2348
\bibitem[\protect\citeauthoryear{Netzel \& Smolec}{2016}]{nspta} Netzel H., Smolec R., 2016, Proceedings of the Polish Astron. Soc., 3, 36 
\bibitem[\protect\citeauthoryear{Netzel, Smolec \& Moskalik}{2015a}]{netzel1} Netzel H., Smolec R., Moskalik P., 2015a, MNRAS, 447, 1173
\bibitem[\protect\citeauthoryear{Netzel, Smolec \& Dziembowski}{2015}]{netzel2} Netzel H., Smolec R., Dziembowski W., 2015, MNRAS, 451, L25
\bibitem[\protect\citeauthoryear{Netzel, Smolec \& Moskalik}{2015b}]{netzel3} Netzel H., Smolec R., Moskalik P., 2015b, MNRAS, 453, 2022
\bibitem[\protect\citeauthoryear{Pietrukowicz et al.}{2013}]{Pietrukowicz2013} Pietrukowicz P., et al., 2013, AcA, 63, 379 
\bibitem[\protect\citeauthoryear{Pietrzy\'nski et al.}{2012}]{bepnat} Pietrzy\'nski et al., 2012, Nature, 484, 75
\bibitem[\protect\citeauthoryear{Poretti}{2001}]{poretti} Poretti E., 2001, A\&A, 371, 986
\bibitem[\protect\citeauthoryear{Simon \& Lee}{1981}]{Simon1981} Simon N.~R., Lee A.~S., 1981, ApJ, 248, 291 
\bibitem[\protect\citeauthoryear{Smolec et al.}{2013}]{bep} Smolec R., et al., 2013, MNRAS, 428, 3034
\bibitem[\protect\citeauthoryear{Smolec et al.}{2015}]{Smolec2015} Smolec R., et al., 2015, MNRAS, 447, 3756 
\bibitem[\protect\citeauthoryear{Smolec et al.}{2016}]{smolec2016} Smolec R., Prudil Z., Skarka M., Bakowska K., 2016, MNRAS, 461, 2934 
\bibitem[\protect\citeauthoryear{Smolec}{2016}]{sm15bl} Smolec R., 2016, Proceedings of the Polish Astron. Soc., 3, 22
\bibitem[\protect\citeauthoryear{Soszy\'nski et al.}{2011}]{ogleIII_rrl_blg} Soszy\'nski I., et al., 2011, Acta Astron., 61, 1
\bibitem[\protect\citeauthoryear{Soszy\'nski et al.}{2014}]{ogleIV_rrl_blg} Soszy\'nski I., et al., 2014, Acta Astron., 64, 177
\bibitem[\protect\citeauthoryear{Soszy\'nski et al.}{2016}]{Soszysky2016MNRAS} Soszy\'nski I., et al., 2016, MNRAS, 463, 1332
\bibitem[\protect\citeauthoryear{Szab{\'o}}{2014}]{Szab2014} Szab{\'o} R., 2014, IAUS, 301, 241 
\bibitem[\protect\citeauthoryear{Szab\'o et al.}{2014}]{szabo_corot} Szab\'o R., Benk\H{o} J.M., Papar\'o M., 2014, A\&A, 570, A100
\bibitem[\protect\citeauthoryear{Udalski et al.}{2008}]{ogleIII} Udalski A., Szyma\'nski M.K., Soszy\'nski I., Poleski R., 2008, Acta Astron., 58, 69
\bibitem[\protect\citeauthoryear{Udalski, Szyma\'nski \& Szyma\'nski}{2015}]{ogleIV} Udalski A., Szyma\'nski M.K., Szyma\'nski G., 2015, Acta Astron., 65, 1
\bibitem[\protect\citeauthoryear{Van Hoolst, Dziembowski \& Kawaler}{1998}]{vdk98} Van Hoolst T., Dziembowski W.A., Kawaler S.D., 1998, MNRAS, 297, 536
\end{thebibliography}




\appendix


\section{Few notes on stars with additional frequencies}
This section is similar to the \ref{subsec:Remarks}. We provide a short list of stars identified as RRab stars, that show additional periodicity but does not fall into our group.

\textbf{OGLE-BLG-RRLYR-08339}. $P_{\mathrm{F}}~=~0.39138132(11)$\,days. Data in OGLE-III and OGLE-IV were available. In addition, we found $P_{\mathrm{A1}}=0.28482069(79)$\,days. The period ratio $P_{\mathrm{A1}}/P_{\mathrm{F}}=0.728$ ($A_{\mathrm{A1}}$/$A_{\mathrm{F}}=0.081$) suggests that the additional frequency corresponds to the first-overtone mode. We found two combination frequencies $\nu_{\mathrm{F}}+\nu_{\mathrm{A1}}$ and $\nu_{\mathrm{A1}}-\nu_{\mathrm{F}}$. Furthermore, we found equidistant peaks in the vicinity of dominant periodicity $k\nu_{\mathrm{F}}+\nu_{\mathrm{B}}$ ($k=1,\ldots, 5$), $k\nu_{\mathrm{F}}-\nu_{\mathrm{B}}$ ($k=1,\ldots, 3$), $k\nu_{\mathrm{F}}+2\nu_{\mathrm{B}}$ ($k=2,\ldots, 3$), $2\nu_{\mathrm{F}}-2\nu_{\mathrm{B}}$ and additional variability $\nu_{\mathrm{A1}}-\nu_{\mathrm{B}}$, $k\nu_{\mathrm{F}}+\nu_{\mathrm{A1}}-\nu_{\mathrm{B}}$ ($k=1,\ldots, 2$), $2\nu_{\mathrm{F}}+\nu_{\mathrm{A1}}+\nu_{\mathrm{B}}$, that most likely represent Blazhko effect with period $P_{\mathrm{B}}=22.496(2)$\,days. In addition, we found one peak for an unknown periodicity $P_{\mathrm{A2}}=0.2634909(34)$\,days. This star is most likely one of the anomalous RRd stars found by \citet[][Galactic bulge]{Smolec2015} and \citet[][Magellanic system]{Soszysky2016MNRAS}.

\textbf{OGLE-BLG-RRLYR-09254}. $P_{\mathrm{F}}~=~0.433295(1)$\,days. Only data in OGLE-III were available. In addition, we found $P_{\mathrm{A}}=0.329295(1)$\,days. The period ratio $P_{\mathrm{A}}/P_{\mathrm{F}}=0.756$ ($A_{\mathrm{A}}/A_{\mathrm{F}}=0.841$) the location in the Petersen diagram (see Fig.~\ref{fig:petersen}) suggests that additional periodicity corresponds to radial first overtone in a HADS star. Furthermore, we found three combination frequencies $\nu_{\mathrm{F}}+\nu_{\mathrm{A}}$, 2$\nu_{\mathrm{F}}+\nu_{\mathrm{A}}$ and $\nu_{\mathrm{A}}-\nu_{\mathrm{F}}$. Most likely F+1O HADS star with long period. 

\textbf{OGLE-BLG-RRLYR-11272}. $P_{\mathrm{F}}~=~0.5086479(1)$\,days. Data in OGLE-III and OGLE-IV were available. In addition, we found $P_{\mathrm{A}}=0.3701556(44)$\,days. The period ratio $P_{\mathrm{A}}/P_{\mathrm{F}}=0.728$ ($A_{\mathrm{A}}$/$A_{\mathrm{F}}=0.157$) suggests that the additional frequency corresponds to the first-overtone mode. We found one combination frequency with the fundamental mode $\nu_{\mathrm{F}}+\nu_{\mathrm{A}}$ and equidistant peaks in the vicinity of $\nu_{\mathrm{F}}$ ($P_{\mathrm{B}}=256.2(7)$\,days) that probably denote to the Blazhko effect. RRd star similar to Blazhko RRd stars with anomalous period ratios found by \cite{Smolec2015} in the Galactic bulge \citep[see also][]{Soszysky2016MNRAS}.

\textbf{OGLE-BLG-RRLYR-11522}. $P_{\mathrm{F}}~=~0.4216513(1)$\,days. Data in OGLE-III and OGLE-IV were available. In addition, we found $P_{\mathrm{A}}=0.2910113(21)$\,days, thus period ratio is $P_{\mathrm{A}}/P_{\mathrm{F}}=0.690$. The additional mode has very low amplitude in comparison with the fundamental mode ($A_{\mathrm{A}}/A_{\mathrm{F}}=0.011$). Due to a low amplitude, the second periodicity was not found in the OGLE-III data. Furthermore, we found three combination frequencies $\nu_{\mathrm{F}}+\nu_{\mathrm{A}}$, 4$\nu_{\mathrm{F}}+\nu_{\mathrm{A}}$ and $\nu_{\mathrm{A}}-\nu_{\mathrm{F}}$. Despite similar properties (period ratio, dominant periodicity) this star does not belong to our newly discovered group. It differs in the shape of the light curve and also this star experience a strong period change.

\textbf{OGLE-BLG-RRLYR-12008}. $P_{\mathrm{F}}~=~0.5560180(2)$\,days. Data in OGLE-III and OGLE-IV were available. In addition, we found $P_{\mathrm{A}}=0.405374(9)$\,days, thus the period ratio is $P_{\mathrm{A}}/P_{\mathrm{F}}=0.729$. The additional mode has very low amplitude in comparison with the fundamental mode ($A_{\mathrm{A}}$/$A_{\mathrm{F}}=0.010$). Furthermore, we found two combination frequencies $\nu_{\mathrm{F}}+\nu_{\mathrm{A}}$ and $\nu_{\mathrm{A}}-\nu_{\mathrm{F}}$. The additional periodicity was also found in OGLE-III data. The Fourier amplitude coeficients $R_{\rm 21}$ and $R_{\rm 31}$ are the highest among RRab stars. This star probably belongs to the group found by \cite{smolec2016}.

\textbf{OGLE-BLG-RRLYR-12069}. $P_{\mathrm{F}}~=~0.382782(34)$\,days. Data in OGLE-III and OGLE-IV were available. In addition, we found $P_{\mathrm{A}}=0.28004(5)$\,days, which yields period ratio $P_{\mathrm{A}}/P_{\mathrm{F}}=0.732$ ($A_{\mathrm{A}}/A_{\mathrm{F}}=0.045$). In the OGLE-IV data we found only one combination freqeuncy $\nu_{\mathrm{F}}+\nu_{\mathrm{A}}$. In the OGLE-III data we did not find any sign of additional varibility. This star is probably another member of anomalous group of RRd stars \citep[see][]{Smolec2015,Soszysky2016MNRAS}, despite its short period and lower amplitude ratio.


\textbf{OGLE-BLG-RRLYR-14965}. $P_{\mathrm{F}}~=~0.33820347(8)$\,days. Only data in OGLE-III were available. In addition, we found $P_{\mathrm{A1}}=0.2563173(5)$\,days. The period ratio $P_{\mathrm{A1}}/P_{\mathrm{F}}=0.758$ ($A_{\mathrm{A1}}/A_{\mathrm{F}}=0.083$) suggests that the additional frequency corresponds to the first-overtone mode. Furthermore, we found three combination frequencies $\nu_{\mathrm{F}}+\nu_{\mathrm{A1}}$, 2$\nu_{\mathrm{F}}+\nu_{\mathrm{A1}}$ and $\nu_{\mathrm{A1}}-\nu_{\mathrm{F}}$. In the close vicinity of the main pulsation frequency and its harmonics we found equidistant peaks that suggests modulation of the fundamental mode with period $P_{\mathrm{B}}=16.46(1)$\,days. We also found one peak for possible third periodicity with period of $P_{\mathrm{A2}}=0.2010675(11)$\,days with period ratio $P_{\mathrm{A2}}/P_{\mathrm{F}}=0.595$. Based on period ratio, it is most likely F+1O HADS star with long period, modulation and possible additional 2O periodicity.




\bsp	
\label{lastpage}
\end{document}